\documentclass[12pt]{article}
\usepackage{amssymb}
\usepackage{epsfig}

\usepackage{graphicx}

\def\L{\mathcal L}
\def\e{\varepsilon}

\newcommand{\wt}{\widetilde}

\begin{document}

\def\a{\alpha}
\def\b{\beta}
\def\c{\chi}
\def\d{\delta}
\def\e{\epsilon}
\def\f{\phi}
\def\g{\gamma}
\def\h{\eta}
\def\i{\iota}
\def\j{\psi}
\def\k{\kappa}
\def\l{\lambda}
\def\m{\mu}
\def\n{\nu}
\def\o{\omega}
\def\p{\pi}
\def\q{\theta}
\def\r{\rho}
\def\s{\sigma}
\def\t{\tau}
\def\u{\upsilon}
\def\x{\xi}
\def\z{\zeta}
\def\D{\Delta}
\def\F{\Phi}
\def\G{\Gamma}
\def\J{\Psi}
\def\L{\Lambda}
\def\O{\Omega}
\def\P{\Pi}
\def\Q{\Theta}
\def\S{\Sigma}
\def\U{\Upsilon}
\def\X{\Xi}

\def\ve{\varepsilon}
\def\vf{\varphi}
\def\vr{\varrho}
\def\vs{\varsigma}
\def\vq{\vartheta}

\def\dg{\dagger}                                     
\def\ddg{\ddagger}                                   
\def\wt#1{\widetilde{#1}}                    
\def\mt{\widetilde{m}_1}
\def\mti{\widetilde{m}_i}
\def\rt{\widetilde{r}_1}
\def\mtt{\widetilde{m}_2}
\def\mttt{\widetilde{m}_3}
\def\rtt{\widetilde{r}_2}
\def\mb{\overline{m}}
\def\VEV#1{\left\langle #1\right\rangle}        
\def\be{\begin{equation}}
\def\ee{\end{equation}}
\def\ds{\displaystyle}
\def\ra{\rightarrow}

\def\bea{\begin{eqnarray}}
\def\eea{\end{eqnarray}}
\def\NO{\nonumber}
\def\Bar#1{\overline{#1}}


\def\pl#1#2#3{Phys.~Lett.~{\bf B {#1}} ({#2}) #3}
\def\np#1#2#3{Nucl.~Phys.~{\bf B {#1}} ({#2}) #3}
\def\prl#1#2#3{Phys.~Rev.~Lett.~{\bf #1} ({#2}) #3}
\def\pr#1#2#3{Phys.~Rev.~{\bf D {#1}} ({#2}) #3}
\def\zp#1#2#3{Z.~Phys.~{\bf C {#1}} ({#2}) #3}
\def\cqg#1#2#3{Class.~and Quantum Grav.~{\bf {#1}} ({#2}) #3}
\def\cmp#1#2#3{Commun.~Math.~Phys.~{\bf {#1}} ({#2}) #3}
\def\jmp#1#2#3{J.~Math.~Phys.~{\bf {#1}} ({#2}) #3}
\def\ap#1#2#3{Ann.~of Phys.~{\bf {#1}} ({#2}) #3}
\def\prep#1#2#3{Phys.~Rep.~{\bf {#1}C} ({#2}) #3}
\def\ptp#1#2#3{Progr.~Theor.~Phys.~{\bf {#1}} ({#2}) #3}
\def\ijmp#1#2#3{Int.~J.~Mod.~Phys.~{\bf A {#1}} ({#2}) #3}
\def\mpl#1#2#3{Mod.~Phys.~Lett.~{\bf A {#1}} ({#2}) #3}
\def\nc#1#2#3{Nuovo Cim.~{\bf {#1}} ({#2}) #3}
\def\ibid#1#2#3{{\it ibid.}~{\bf {#1}} ({#2}) #3}

\title{
\vspace*{15mm}
\bf Flavor effects on leptogenesis predictions}
\author{{\Large Steve Blanchet and Pasquale Di Bari}
\\
{\it Max-Planck-Institut f\"{u}r Physik} \\
{\it (Werner-Heisenberg-Institut)} \\
{\it F\"{o}hringer Ring 6, 80805 M\"{u}nchen}}

\maketitle \thispagestyle{empty}


\begin{abstract}
We show that flavor effects in leptogenesis reduce
the region of the see-saw parameter space where the
final predictions do not depend on the initial conditions,
the strong wash-out regime. In this case
the lowest bounds holding on the lightest right-handed (RH)
neutrino mass and on the reheating temperature for hierarchical
heavy neutrinos do not get relaxed compared to the usual ones in
the one-flavor approximation,
$M_1 (T_{\rm reh})\gtrsim 3\,(1.5)\times 10^{9}\,{\rm GeV}$.
Flavor effects can however relax down to these minimal values
the lower bounds holding for fixed large values of the decay parameter
$K_1$. We discuss a relevant definite example showing
that, when the known information on the neutrino mixing matrix
is employed, the lower bounds for $K_1\gg 10$ are relaxed by a
factor 2-3 for fully hierarchical light neutrinos, without any
dependence on $\theta_{13}$ and on possible phases.
On the other hand, going beyond the limit of hierarchical light
neutrinos and taking into account Majorana phases,
the lower bounds can be relaxed by one order of magnitude.
Therefore, Majorana phases can play an important role
in leptogenesis when flavor effects are included.
\end{abstract}

\newpage
\section{Introduction}

The see-saw mechanism \cite{seesaw} is a simple and
elegant way to understand neutrino masses and their lightness
compared to all other fermions. At the same time,
with  leptogenesis \cite{fy}, the see-saw
offers an attractive model to explain the
matter-antimatter asymmetry of the Universe. The asymmetry
is produced in the decays of the very heavy RH neutrinos,
necessary for the see-saw mechanism to work, generating a $B-L$ asymmetry
in the form of a lepton asymmetry  partly converted into a baryon
asymmetry by sphaleron processes
if temperatures are larger than about $100\,{\rm GeV}$ \cite{sphalerons}.

Calculations have been typically performed within a one-flavor
approximation, assuming that the final total $B-L$
asymmetry is not sensitive to the dynamics of the
individual $B/3-L_{\a}$ asymmetries.
In this case it has been noticed \cite{bdp2,annals}
that current neutrino mixing data favor leptogenesis to lie in
a region of the parameter space where thermal equilibrium sets up
prior to the freeze-out of the final asymmetry and therefore, if the
temperature of the early Universe is large enough, the final asymmetry
does not depend on the initial conditions.

For a hierarchical heavy neutrino spectrum, successful leptogenesis
requires the mass of the lightest \cite{di} or of the
next-to-lightest \cite{geometry} RH neutrino to be large enough,
if the RH neutrino production occurs through scatterings in the
thermal plasma. In this case there is also an associated lower bound
on the initial temperature of the radiation-dominated regime in the
early Universe, to be identified with the reheating temperature,
$T_{\rm reh}$, within inflation. Being  independent of the initial conditions,
these bounds are particularly relevant, since they
appear as an intrinsic feature of leptogenesis.

Recently it has been pointed out that flavor effects
can strongly modify leptogenesis predictions
\cite{bcst,endo,pilaftsis,vives,nardi,abada,abada2}. In this work
we investigate how the dependence on the initial conditions and the lower bounds
on the RH neutrino masses and on the reheating temperature are
modified when flavor effects are included. We show that the usually
quoted values do not get relaxed, contrarily to what has been stated
in \cite{abada}.
The calculations are presented within the hierarchical limit for
the spectrum of the heavy neutrino masses, and more specifically in the
lightest RH neutrino dominated scenario ($N_1$DS), where the
final asymmetry is dominantly produced by the lightest
RH neutrino decays. In the second Section we set up the notation and
the kinetic equations. In the third Section we analyze
how flavor effects modify the dependence on the initial conditions.
In the fourth Section
we perform explicit calculations for a specific but significant
example. In the Appendix we discuss the role of $\D L=1$ scatterings, showing
how they introduce just a correction to the results.

\section{General set-up}

Adding to the Standard Model Lagrangian three RH neutrinos with a
Yukawa coupling matrix $h$ and a Majorana mass matrix $M$, a neutrino
Dirac mass matrix $m_D=h\,v$ is generated, after electro-weak symmetry
breaking, by the vev $v$ of the Higgs boson. For $M\gg m_D$, the
neutrino mass spectrum splits into 3 heavy Majorana states $N_1$,
$N_2$ and $N_3$ with masses $M_1\leq M_2\leq M_3$, almost
coinciding with the eigenvalues of $M$, and 3 light Majorana states
with masses $m_1\leq m_2 \leq m_3$, corresponding to the
eigenvalues of the neutrino mass matrix given by the see-saw formula,
\be\label{seesaw}
m_{\nu}=-m_D\,{1\over M}\,m_D^T \, .
\ee
Neutrino mixing experiments measure two mass-squared differences.
In normal (inverted) neutrino schemes, one has
\be
m^2_3-m^2_2= \D m^2_{\rm atm}\,\,(\D m_{\rm sol}^2) \,\, ,
\ee
\be
m^2_2-m^2_1= \D m^2_{\rm sol}
\,\,(\D m_{\rm atm}^2-\D m^2_{\rm sol}) \, .
\ee
For
$m_1\gg m_{\rm atm}\equiv
\sqrt{\D m_{\rm atm}^2+\D m_{\rm sol}^2}
\simeq 0.05\,{\rm eV}$,
one has a quasi-degenerate
spectrum with $m_1\simeq m_2 \simeq m_3$, whereas for
$m_1\ll m_{\rm sol}\equiv \sqrt{\D m^2_{\rm sol}}\simeq 0.009\,{\rm eV}$
one has a fully hierarchical (normal or inverted) spectrum.

 A lepton asymmetry can be generated from the decays of the
heavy neutrinos into leptons and Higgs bosons and
partly converted into a baryon asymmetry by the sphaleron
($B-L$ conserving) processes at temperatures higher
than about $100\,{\rm GeV}$.

For $10^{9}\,{\rm GeV}\lesssim M_1 \lesssim 10^{12}\,{\rm GeV}$,
tauon charged lepton Yukawa interactions are in
equilibrium~\cite{bcst,abada,nardi} and typically
\footnote{See also the note added in the end of the paper.}
faster than inverse processes \cite{zeno},
breaking the coherent evolution of the lepton doublets
quantum states, the $|l_i \rangle$'s, produced in RH neutrino decays.
The quantum state is projected on a two-flavor basis, where the eigenstates
are given by the tauon flavor and by a linear combination of muon
and electron flavors.
For $M_1\lesssim 10^9\,{\rm GeV}$ both muon and
tauon charged lepton Yukawa interactions are strong
enough that the quantum state is projected on a
three-flavor basis.

Flavor effects play a double role in leptogenesis.
A first effect is that the wash-out is reduced. This happens
because, in the inverse decays, the Higgs do not interact
with $|l_i\rangle$'s but with the flavor eigenstates
$|l_{\alpha}\rangle$'s, with a reduced inverse decay rate.
This effect can be accounted for introducing the flavor projectors
\cite{bcst,nardi}
\bea
\hspace{37mm}
P_{i\a} & = & |\langle l_i|l_{\a}\rangle|^2  =
{\G_{i\a}\over\G_i}\, , \\
\overline{P}_{i\a} & = & |\langle \bar{l}_i'|\overline{l}_{\a}\rangle|^2
 =  {\overline{\G}_{i\a}\over\overline{\G}_i}\, ,\hspace{15mm}
(i=1,2,3; \a=e,\m,\t)
\eea
where $\G_{i\a}$ is the partial decay rate of the process
$N_i\rightarrow l_{\a}+H^{\dagger}$ and $\G_i=\sum_{\a}\,\G_{i\a}$
is the total decay rate, such that
$\sum_{\a}\,P_{i\a}=1$. Analogously $\overline{\G}_{i\a}$
is the partial decay rate of the process $N_i\rightarrow \bar{l}_{\a}+H$
with $\overline{\G}_i=\sum_{\a}\,\overline{\G}_{i\a}$,
such that $\sum_{\a}\,\overline{P}_{i\a}=1$.

A rigorous description of the asymmetry
evolution has then to be performed in terms of the
individual flavor asymmetries $\D_{\a}\equiv B/3-L_{\a}$
rather than in terms of the total asymmetry
$N_{B-L}=\sum_{\a}\,N_{\D_{\a}}$,
as usually done in the one-flavor approximation.
The contribution of each decay to the $N_{\D_{\a}}$'s
is determined by the individual flavor $C\!P$ asymmetry, defined as
\be
\ve_{i\a}\equiv -{\G_{i\alpha}-\overline{\G}_{i\alpha}
\over \G_{i}+\overline{\G}_{i}} \, .
\ee
A second role played by flavor effects arises because
the state $|\bar{l}_i'\rangle$ is not the
$C\!P$ conjugated state of $|l_i\rangle$  \cite{bcst,nardi}
and this yields an additional source of $C\!P$ violation.
This can be described in terms of the projector differences \cite{bcst,nardi}
\be
\D P_{i\a}\equiv P_{i\a}-\bar{P}_{i\a} \, ,
\ee
obeying $\sum_{\a}\,\D P_{i\a}=0$.
Indeed writing $P_{i\a}=P^0_{i\a}+\D P_{i\a}/2$ and
$\overline{P}_{i\a}=P^0_{i\a}-\D P_{i\a}/2$, where the
$P^0_{i\a}\equiv (P_{i\a}+\overline{P}_{i\a})/2$ are
the tree level contributions to the projectors, one can
see that
\be
\ve_{i\a}=\ve_i\,P^{0}_{i\a}+ {\D\,P_{i\a}\over 2} \, ,
\ee
where the total $C\!P$ asymmetries $\ve_i\equiv \sum_{\a}\,\ve_{i\a}$.
The three flavored $C\!P$ asymmetries can be calculated using \cite{crv}
\be\label{eps1a}
\ve_{i\a}=
\frac{3}{16 \p (h^{\dag}h)_{ii}} \sum_{j\neq i} \left\{ {\rm Im}\left[h_{\a i}^{\star}
h_{\a j}(h^{\dag}h)_{i j}\right] \frac{\x(x_j/x_i)}{\sqrt{x_j/x_i}}+
\frac{2}{3(x_j/x_i-1)}{\rm Im}
\left[h_{\a i}^{\star}h_{\a j}(h^{\dag}h)_{j i}\right]\right\} \, ,
\ee
where $x_i\equiv (M_i/M_1)^2$ and
\be\label{xi}
\xi(x)= {2\over 3}\,x\,
\left[(1+x)\,\ln\left({1+x\over x}\right)-{2-x\over 1-x}\right] \, .
\ee
In the last years there has been an intense study of the
relevant processes in leptogenesis \cite{many,bcst,many2,giudice,spectator}.
For our purposes it will be enough to describe
the evolution of the asymmetries just in terms of
decays and inverse decays (after subtraction of the
real intermediate state contribution from $\D L=2$ processes)
neglecting $\D L=1$ scatterings
and thermal effects. In the Appendix we will discuss and
justify this approximation. We will also neglect off-shell
$\D L=2$ and $\D L=0$ processes, relevant
only at higher temperatures, and spectator processes
that produce corrections not larger than $30\%$, within
the precision needed for our purposes \cite{spectator}.
The kinetic equations, using
$z\equiv M_1/T$ as the independent variable, are then given by
\cite{bcst,nardi,abada,blanchet}
\bea\label{flke}
{dN_{N_i}\over dz} & = & -D_i\,(N_{N_i}-N_{N_i}^{\rm eq}) \\ \nonumber
{dN_{\D_{\a}}\over dz} & = &
\sum_i\,\ve_{i\a}\,D_i\,(N_{N_i}-N_{N_i}^{\rm eq})
-N_{\D_{\a}} \,\sum_i\,P_{i\a}^{0}\,W_i^{\rm ID}\, ,
\eea
where the $N_{\D_{\a}}$'s and the ${N_i}$'s are the abundances
per number of $N_1$'s in ultra-relativistic
thermal equilibrium. Notice that $\a=\e,\m,\t$ in the three-flavor case,
while $\a=\t,e+\m$ in the two-flavor case, where `$e+\m$' means that
the electron and muon  $C\!P$
asymmetries $\ve_{i\a}$ and projectors $P_{i\a}^0$, are summed.
Therefore, in the two-flavor case, there are only two kinetic
equations for the $N_{\D_{\a}}$ instead of three.
The equilibrium abundances are given by
$N_{N_i}^{\rm eq}=z_i^2\,{\cal K}_2(z_i)/2$, where we indicate with
${\cal K}_i(z_i)$ the modified Bessel functions.
Introducing the decay parameters
$K_i\equiv \G_{D,i}(T=0)/H(T=M_i)$,
the ratios of the total decay widths to the
expansion rate at $T=M_i$,
the decay terms $D_i$ can be written like \cite{annals}
\be
D_i \equiv {\G_{D,i}\over H\,z}=K_i\,x_i\,z\,
\left\langle {1\over\gamma_i} \right\rangle   \, ,
\ee
where $\G_{D,i}\equiv \G_{i}+\overline{\G}_{i}
=\G_{D,i}(T=0)\,\langle 1/\gamma_i\rangle$ are the total
decay rates and the $\langle 1/\gamma_i\rangle$'s are
the thermally averaged dilation factors and are given by the
ratios ${\cal K}_1(z_i)/{\cal K}_2(z_i)$. Finally,
the inverse decays wash-out terms are given by
\be\label{WID}
W_i^{\rm ID}(z) =
{1\over 4}\,K_i\,\sqrt{x_i}\,{\cal K}_1(z_i)\,z_i^3 \, .
\ee
The evolution of the $N_{\D\a}$'s can be worked out in an integral form,
\be
N_{\D\a}(z)=N_{\D\a}^{\rm in}\,
e^{-\sum_i\,P_{i\a}^0\,\int_{z_{\rm in}}^z\,dz'\,W_i^{\rm ID}(z')}
+\sum_i\,\ve_{i\a}\,\k_{i{\a}}(z) \,  ,
\ee
with  the 6, in the two-flavor case, or 9, in the
three-flavor case, efficiency factors given by
\be\label{ef}
\k_{i\a}(z;K_i,P^{0}_{i\a})=-\int_{z_{\rm in}}^z\,dz'\,{dN_{N_i}\over dz'}\,
e^{-\sum_i\,P_{i\a}^0\,\int_{z'}^z\,dz''\,W_i^{\rm ID}(z'')} \,.
\ee
The total final $B-L$ asymmetry is then given by
$N_{B-L}^{\rm f}=\sum_{\a}\,N_{\D_\a}^{\rm f}$. Finally from this,
assuming a standard thermal history and accounting for the
sphaleron converting coefficient $a_{\rm sph}\sim 1/3$,
the final baryon-to-photon number ratio can be calculated as
\be\label{etaB}
\eta_B=a_{\rm sph}\,{N_{B-L}^{\rm f}\over N_{\gamma}^{\rm rec}}
\simeq 0.96\times 10^{-2}\,N_{B-L}^{\rm f} \, ,
\ee
to be compared with the measured value \cite{WMAPSLOAN}
\be\label{etaBobs}
\eta_B^{\rm CMB} = (6.3 \pm 0.3)\times 10^{-10} \, .
\ee
Notice that the efficiency factors depend only on the
$P_{i\a}^0$ but not on the differences $\D P_{i\a}$.
The final asymmetry depends in general on all the unmeasured 14
see-saw parameters. A useful parametrization
is provided by the orthogonal see-saw matrix $\O$ \cite{casas},
in terms  of which one has $m_D=U\,\sqrt{D_m}\,\O\,\sqrt{D_M}$.
In this way the decay parameters can be expressed as
\be
K_i=\sum_j\,{m_j\over m_{\star}}\,|\O_{ji}^2| \, ,
\ee
where $m_{\star}\simeq 10^{-3}\,{\rm eV}$ is the
equilibrium neutrino mass, and the tree level projectors as
\be\label{proj}
P^0_{i\a}={|\sum_j\,\sqrt{m_j}\,U_{\a j}\,\O_{j i}|^2
\over \sum_j\,m_j\,|\O^2_{ji}|} \, .
\ee
It will also prove useful to introduce the quantities
$K_{i\a}\equiv P_{i\a}^{0}\,K_i$.
The orthogonal matrix  can in turn be decomposed as
\be
\O=R_{12}(\o_{21})\,R_{13}(\o_{31})\,R_{23}(\o_{22})\, ,
\ee
such to be parameterized in terms of three complex numbers,
$\o_{21}$, $\o_{31}$ and $\o_{22}$, determining
the rotations in the planes 12, 13 and 23 respectively.
In this way one has $\eta_B=\eta_B(m_1,U,M_i,\o_{ij})$.
It is interesting that including flavor effects there is a
potential dependence
of the final asymmetry also on the unknown parameters contained
in the PMNS mixing matrix $U$ \cite{endo,nardi}.

In the following we will assume a hierarchical heavy neutrino
spectrum with $M_3,M_2\gg M_1$
\footnote{Just in passing, we notice that it is
straightforward to generalize, including flavor effects,
a result obtained in \cite{blanchet}
for the efficiency factors in the degenerate limit, obtained for
$(M_3-M_1)/M_1 \lesssim 0.1$, within the one-flavor approximation and
for $K_i \gg 1$, for all $i$. Indeed, in this case one can approximate
$dN_i/dz'\simeq dN_i^{\rm eq}/dz'$ in the Eq. (\ref{ef}) obtaining
that $\k_{i\a}^{\rm f}\simeq \k(K_{\a})$, where
$K_{\a}\equiv \sum_{i}\,K_{i\a}$. The function
$\k(x)$ is defined in the  Eq. (\ref{k1a}) and it
approximates $\k_{1\a}^{\rm f}$ in the hierarchical limit when
$x=K_{1\a}$. In the degenerate limit one has then just to replace
$K_{1\a}$ with the sum $K_{\a}$. Like in the hierarchical limit, the number of
efficiency factors to be calculated reduces from 6 to 2 in the two-flavor
case and from 9 to 3 in the three-flavor case, one for each flavor.
If instead of a full degeneracy, one has just a partial degeneracy,
with $M_3\gg M_2\simeq M_1$, then simply one has
$\k_{3\a}^{\rm f}\ll \k_{1\a}^{\rm f}\simeq \k_{2\a}^{\rm f}
\simeq \k(K_{1\a}+K_{2\a})$.}.
In this case the general expression Eq. (\ref{eps1a}) for
the $C\!P$ asymmetries $\ve_{1\a}$, reduces to
\be
\ve_{1\a}={1\over 8\,\pi\,(h^{\dagger}\,h)_{11}}
\,\sum_{j\neq 1}\,
\left\{{\rm Im}\left[h^{\star}_{\a 1}\,h_{\a j}
\left({3\over 2\,\sqrt{x_j}}(h^{\dagger}h)_{1j}
+{1\over x_j}(h^{\dagger}h)_{j1}\right)\right] \right\} \, .
\ee
We will moreover assume no rotation in the plane 23, i.e. $R_{23}=1$.
Under these conditions
both the total $C\!P$ asymmetries $\ve_2$ and $\ve_3$
are suppressed $\sim M_1/M_{2,3}$. It is interesting to notice
that, under particular conditions, the $\ve_{2\a}$'s and the
$\ve_{3\a}$'s  are not suppressed and do not vanish.
This can potentially lead to a scenario where the
final asymmetry is produced by the decays of the two heavier
RH neutrinos if $M_{2,3}\lesssim 10^{12}\,{\rm GeV}$
such that the flavored regime applies.
Here we do not pursue this possibility and
focus on a typical $N_1$-dominated scenario where
the dominant contribution to the final asymmetry comes
from the decays of the lightest RH neutrino and
\be\label{NBmL}
N_{B-L}^{\rm f}\simeq\,\sum_{\a}\,\ve_{1\a}\,\k_{1\a}^{\rm f} \, .
\ee
It will prove important for our discussion that the
$C\!P$ asymmetries, the total $\ve_1$ and the flavored $\ve_{1\a}$'s,
cannot be arbitrarily large. The
total $C\!P$ asymmetry is indeed upper bounded by \cite{asaka,di,bdp2}
\be\label{boundtot}
\ve_1 \leq \overline{\ve}(M_1)\,\b_{\rm max}\,f(m_1,K_1) ,
\ee
where $\overline{\ve}(M_1) \equiv 3\,M_1\,m_{\rm atm}/(16\,\pi\,v^2)$,
$\b_{\rm max}\equiv m_{\rm atm}/(m_1+m_3)$ and
$0\leq f(m_1,K_1)\leq 1$
is a function that vanishes for $K_1=m_1/m_{\star}$ and tends to
unity for $m_1/(m_{\star}\,K_1)\rightarrow 0$. On the other hand
each individual flavor asymmetry $\ve_{1\a}$ is bounded by
\cite{abada2}
\be\label{bound}
|\ve_{1\a}|< \overline{\ve}(M_1)\,\sqrt{P_{1\a}^0}\,
{m_3\over m_{\rm atm}}\,
{\rm max}_{\rm j}\,\,[|U_{\a j}|] \,
\ee
and therefore while the total $C\!P$ asymmetry is suppressed when
$m_1$ increases, the single flavor asymmetries can be enhanced.
The existence of an upper bound on the quantity
$r_{1\a}\equiv \ve_{1\a}/\overline{\ve}(M_1)$ independent of $M_1$,
imply, as in the case of one-flavor approximation,
the existence of a lower bound on $M_1$ given by
\be\label{lbM1}
M_1  \geq M_1^{\rm min}\simeq
        {N_{\g}^{\rm rec}\over a_{\rm sph}}
           \,{16\,\pi\,v^2\over 3}\,
          {\eta_B^{\rm CMB}/m_{\rm atm}\over
          \k^{\rm f}_1(K_1)\,\xi_1(K_1)}
          \geq
{4.2\times 10^{8}\,{\rm GeV} \over \k_1^{\rm f}(K_1)\,\xi_1(K_1)}
\hspace{3mm}(\mbox{at}\,\,3\,\s\,\,\mbox{C.L.} ) \, ,
\ee
where we indicated with $\k^{\rm f}_1(K_1)$ the efficiency factor
in the usual one-flavor approximation, corresponding to
$\k_{1\a}^{\rm f}$ with $P^0_{1\a}=1$. We also defined
\be\label{xi1a}
\xi_1 \equiv \sum_{\a}\,\xi_{1\a}\, ,
\hspace{10mm}\mbox{with}\hspace{12mm}
\xi_{1\a}\equiv {r_{1\a}\,\k_{1\a}^{\rm f}(K_{1\a})
                      \over \k_1^{\rm f}(K_1)}\, .
\ee
This quantity represents the deviation introduced by flavor
effects compared to the one-flavor approximation in the
hierarchical light neutrino case.
We could then use $r_{1\a}\leq\sqrt{P^0_{1\a}}\,m_3/m_{\rm atm}$
to maximize $\xi_1$.
Notice however that first, because of the bound on the
total asymmetry, the $r_{1\a}$'s cannot
be simultaneously equal to $\sqrt{P^0_{1\a}}$ and
second in $\xi_1$ there are sign cancellations.
Therefore, the bound Eq. (\ref{lbM1}) is more restrictive
than this possible estimation and we prefer to keep it in this form,
specifying $\xi_1$ in particular situations.

As usual, the lower bound on $M_1$ imply also an associated lower bound
on the reheating temperature $T_{\rm reh}$.

\section{Dependence on the initial conditions}

From the Eq. (\ref{ef}), extending an analytic procedure
derived within the one-flavor approximation \cite{annals},
one can obtain simple expressions for the $\k_{1\a}^{\rm f}$'s.
In the case of an initial thermal abundance
($N_{N_1}^{\rm in}=1$) one has
\be\label{k1a}
\k_{1\a}^{\rm f} \simeq \k(K_{1\a}) \equiv
{2\over K_{1\a}\,z_B(K_{1\a})}\,
\left(1-e^{-{K_{1\a}\,z_B(K_{1\a})\over 2}}\right) \, ,
\ee
where
\be
z_{B}(K_{1\a}) \simeq 2+4\,K_{1\a}^{0.13}\,e^{-{2.5\over K_{1\a}}} \, .
\ee
Notice that in the particularly relevant range
$5\lesssim K_{1\a}\lesssim 100$ this expression is well
approximated by a power law \cite{newsvenezia},
\be\label{plaw}
\k_{1\a}^{\rm f}\simeq {0.5\over K_{1\a}^{1.2}} \, .
\ee
In the case of initial vanishing abundance ($N_{N_1}^{\rm in}=0$)
one has to take into account two different contributions,
a negative and a positive one,
\be
\k_{1\a}^{\rm f}
=\k_{-}^{\rm f}(K_1,P_{1\a}^{0})+
 \k_{+}^{\rm f}(K_1,P_{1\a}^{0}) \, .
\ee
The negative contribution arises from a first stage when
$N_{N_1}\leq N_{N_1}^{\rm eq}$, for $z\leq z_{\rm eq}$,
and is given approximately by
\be\label{k-}
\k_{-}^{\rm f}(K_1,P_{1\a}^{0})\simeq
-{2\over P_{1\a}^{0}}\ e^{-{3\,\pi\,K_{1\a} \over 8}}
\left(e^{{P_{1\a}^{0}\over 2}\,N_{N_1}(z_{\rm eq})} - 1 \right) \, .
\ee
The $N_1$ abundance at $z_{\rm eq}$ is well approximated by
\begin{equation}\label{nka}
N_{N_1}(z_{\rm eq}) \simeq \overline{N}(K)\equiv
{N(K_1)\over\left(1 + \sqrt{N(K_1)}\right)^2}\, ,
\end{equation}
interpolating between the limit $K_1\gg 1$, where $z_{\rm eq}\ll 1$ and
$N_{\rm N_1}(z_{\rm eq})=1$, and the limit $K_1\ll 1$, where
$z_{\rm eq}\gg 1$ and $N_{N_1}(z_{\rm eq})=N(K_1)\equiv 3\p K_1/4$.
The positive contribution arises from a second stage when
$N_{N_1}\geq N_{N_1}^{\rm eq}$, for $z\geq z_{\rm eq}$,
and is approximately given by
\be\label{k+}
\k_{+}^{\rm f}(K_1,P_{1\a}^{0})\simeq
{2\over z_B(K_{1\a})\,K_{1\a}}
\left(1-e^{-{K_{1\a}\,z_B(K_{1\a})\,N_{N_1}(z_{\rm eq})\over 2}}\right) \, .
\ee
Notice, from the Eq. (\ref{nka}),
how $N_{N_1}(z_{\rm eq})$ is still regulated by $K_1$ and
this because the RH neutrino production is not affected by flavor
effects, contrarily to the wash-out, which is reduced
and is regulated by $K_{1\a}$.

These analytic expressions make transparent the two conditions
to have independence on the initial conditions.
The first is the thermalization of the $N_1$ abundance,
such that, for an arbitrary initial abundance
one has $N_{N_1}(z_{\rm eq})=1$.
The second is that the asymmetry produced
during the non-thermal stage, for $z\leq z_{\rm eq}$,
has to be efficiently washed out
such that $|\k_{-}^{\rm f}| \ll \k_{+}^{\rm f}$. They
are both realized for large values of $K_1\gg 1$.
More quantitatively, it is useful to introduce
a value  $K_{\star}$ such that, for
$K_1\geq K_{\star}$, the final asymmetry calculated
for initial thermal abundance differs from that one
calculated for vanishing initial abundance by less than some quantity
$\d$. This can be used as a precise definition of
the strong wash-out regime.
Let us consider some particular cases, showing
how flavor effects tend to enlarge the domain of the
weak wash-out at the expense of the strong wash-out regime.

\subsection{Alignment}

The simplest situation is the alignment case,
realized when the $N_1$'s decay  just into one-flavor
$\a$, such that $P_{1\a}=\overline{P}_{1\a}=1$ and
$P_{1\b\neq\a}=\overline{P}_{1\b\neq\a}=0$, implying
$\ve_{1\a}=\ve_1$. Notice that we do not have to worry about
the fact that
the lightest RH neutrino inverse decays might not be able
to wash-out the asymmetry generated from the decays of the
two heavier ones, since we are assuming negligible
$\ve_{2\b}$ and $\ve_{3\b}$ anyway.
In this case the general set of kinetic equations (\ref{flke})
reduces to the usual one-flavor case and all results coincide
with those in the one-flavor approximation \cite{nardi}.
In particular one  has $N_{B-L}^{\rm f}=\ve_1\,\k_{1\a}^{\rm f}$.

In the considered case of alignment, one obtains
$K_{\star}\simeq 3.5$ for $\d=0.1$, as shown in Fig. 1.
\begin{figure}
\hspace*{-5mm}
\psfig{file=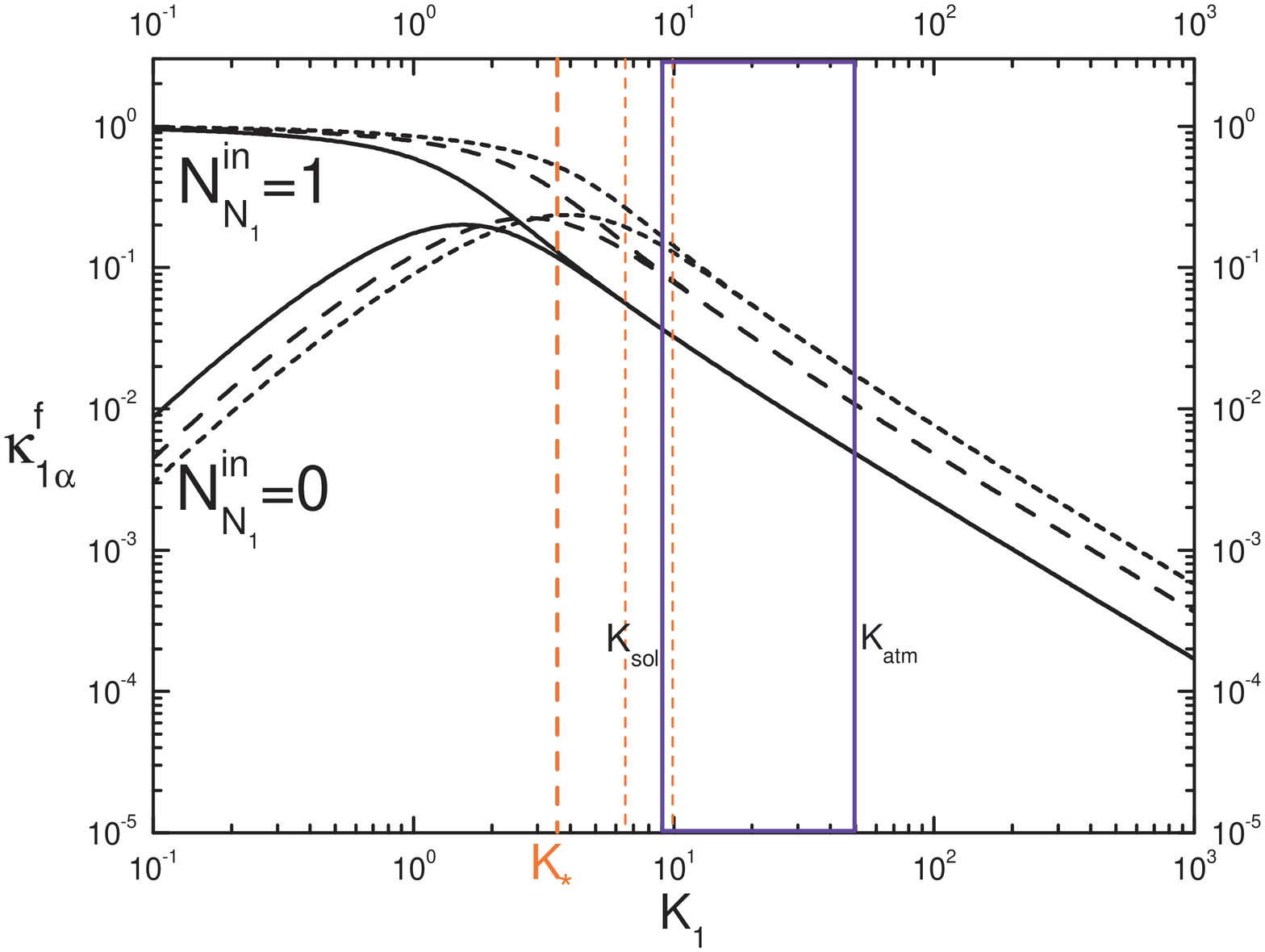,height=8cm,width=9cm}
\hspace{-10mm}
\psfig{file=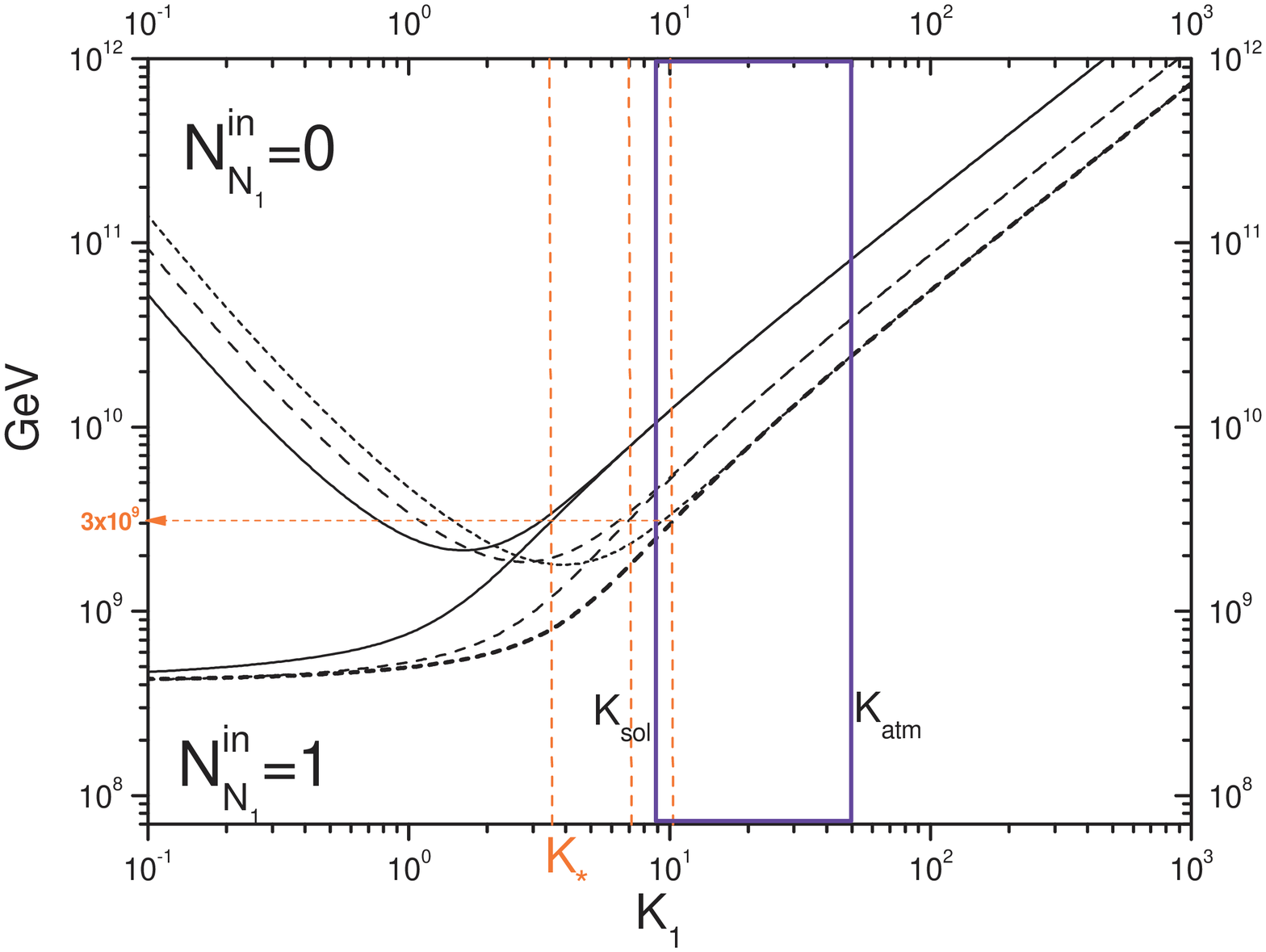,height=8cm,width=9cm}
\caption{Efficiency factor (left panel) and lower bounds on
$M_1$  (right panel) in the alignment (solid),
in the semi-democratic (dashed lines)
and in the democratic (short-dashed) cases.}
\end{figure}
The value of $K_{\star}$ plays a  relevant role since
only for $K_1\gtrsim K_{\star}$ one has  leptogenesis
predictions on the final baryon asymmetry resulting from
a self-contained set of assumptions. On the other hand
for $K_1\lesssim K_{\star}$ leptogenesis has to be complemented
with a model for the initial conditions.

In addition to that, one has to say that in the weak wash-out regime
the calculation of the final asymmetry requires a precise description of
the RH neutrino production, potentially sensitive to many poorly known
effects. It is then interesting that
current neutrino mixing data favor $K_1$ to be in the range
$K_{\rm sol}\simeq 9 \lesssim K_1 \lesssim 50 \simeq K_{\rm atm}$
\cite{bdp2,annals,geometry}, where one can have a mild wash-out
assuring full independence of the initial condition, as one can
see in Fig. 1, but still successful leptogenesis.
In this case one can place constraints
on the see-saw parameters not depending on specific assumptions
for the initial conditions and with reduced theoretical uncertainties.

Since one has $\xi_1=1$, the general lower bound on $M_1$
(see Eq. (\ref{lbM1})), like all other quantities,
becomes the usual lower bound holding in the
one-flavor approximation.
In the right panel of Fig. 1 we have plotted it
both for $N_{N_1}^{\rm in}=1$ and $N_{N_1}^{\rm in}=0$. One can see
that the $\k_{1\a}^{\rm f}$ dependence on the initial conditions
translates into a dependence of $M_1^{\rm min}$.
The lowest model-independent values are then obtained
for $K_1= K_{\star} \simeq 3.5$ and are given by
\be
M_1\gtrsim 3\times 10^{9}\,{\rm GeV}
\hspace{10mm}\mbox{and}\hspace{10mm}
T_{\rm reh}\gtrsim 1.5\times 10^{9}\,{\rm GeV}\, .
\ee
Another typically quoted lower bound on $M_1$
is obtained for initial thermal abundance
in the limit $K_1\rightarrow 0$, given
by $M_1\gtrsim 4\times 10^8\,{\rm GeV}$ \cite{bdp}.

\subsection{Democratic and semi-democratic cases}

Let us now discuss another possibility and for definiteness
let us first consider the three-flavor regime. We will easily
extend the results to the two-flavor regime.
Let us assume that $P_{1\a}=\bar{P}_{1\a}=1/3$ for
any $\a$ and consequently $\D\,P_{1\a}=0$.
This case was also considered in \cite{abada}.
From the
Eq.'s (\ref{ef}) it follows that the three
$\k_{1\a}^{\rm f}$, like also the three $\ve_{1\alpha}$,
are all equal and thus the Eq. (\ref{NBmL}) simplifies into
$N_{B-L}^{\rm f}=\ve_1\,\k_{1\,\a}^{\rm f}$,
as in the usual one-flavor approximation.
However now the wash-out is reduced by the
presence of the projector, such that
$K_1\rightarrow P^{0}_{1\a}\,K_1=K_1/3$.
The result is that in the case of an initial thermal abundance
the efficiency factor, as a function of $K_1$, is simply shifted.
The same happens for vanishing initial abundance in the strong
wash-out regime. However, in the weak wash-out regime,
there is not a simple shift,
since the RH neutrino production is still depending on $K_1$.
A plot of $\k_{1\a}^{\rm f}$ is shown in the left panel of Fig. 1
(dashed lines).
One can see how the reduced wash-out increases
the value of $K_{\star}$ to $\sim 10$, approximately
$1/P^0_{1\a}\simeq 3$ larger, thus compensating almost
exactly the wash-out reduction by
a factor $\sim 3^{1.2}$ (cf. (\ref{plaw})).

In this way the lowest bounds on $M_1$ in
the strong wash-out regime, at $K_1=K_{\star}$,
are almost unchanged. On the other hand
for a given value of $K_1\gg K_{\star}$, the lower bounds get approximately
relaxed by a factor $3$ \cite{abada}. The lower bounds for
the democratic case are shown in the right panel of Fig. 1 (dashed lines).
Notice that also the lower bound at $K_1\rightarrow 0$ for
initial thermal abundance does not change. More generally,
flavor effects simply induce a shift
of the dependence of the lower bound on $K_1$.

The semi-democratic case is intermediate
between the democratic and the alignment cases. It
is obtained when one projector vanishes, for example $P_{1\b}=0$, and
the other two are one half and in this case  $K_{\star}\sim 7$.
The corresponding plot of the efficiency factor and of the lower bound
on $M_1$ are also shown in Fig. 1 (short-dashed lines). These results
for the semi-democratic case also apply to the two-flavor regime,
when the two projectors are equal, namely $P^0_{\t}=P^0_{e+\mu}=1/2$.

\subsection{One-flavor dominance}

There is another potentially interesting situation that motivates
an extension of the previous results to arbitrarily
small values of $P_{1\a}^{0}$. This occurs when
the final asymmetry is dominated by one flavor $\a$
and the Eq. (\ref{NBmL}) can be further simplified into
\be
N_{B-L}^{\rm f}\simeq \ve_{1\a}\,\k_{1\a}^{\rm f} \, ,
\ee
analogously to the alignment case but with $P^0_{1\a}\ll 1$.
Notice that this cannot happen due to a dominance of one of the
$C\!P$ asymmetries, for example with $\ve_{1\a}$ being close to
its maximum value, Eq. (\ref{bound}), much larger than the other two
strongly suppressed, simply because one has $\sum_{\a}\,\D P_{1\a}=0$.
One has then to imagine a situation where the $C\!P$ asymmetry
$\ve_{1\a}$ is comparable to the sum of the other two but
$K_{1\b\neq\a}\gg K_{1\a}\gtrsim 1$ such that
$\k_{1\a}^{\rm f}\gg \k_{1\b}^{\rm f}$. The dominance
is then a result of the much weaker wash-out.

The analysis of the dependence on the initial conditions can then be again
performed as in the previous cases calculating the value of $K_{\star}$ for
any value of $P_{1\a}^{0}$. The result is shown in Fig. 2.
 The alignment case corresponds to $P_{1\a}^{0}=1$, the semi-democratic
 case to $P_{1\a}^{0}=1/2$ and the democratic case to $P_{1\a}^{0}=1/3$.
 Notice that the result is very close to the simple estimation
 $K_\star(P^0_{1\a})=K_\star(1)/P^0_{1\a}$,
that would follow if $\k_{1\a}^{\rm f}$ were just depending on $K_{1\a}$.
\begin{figure}
\hspace*{25mm}
\psfig{file=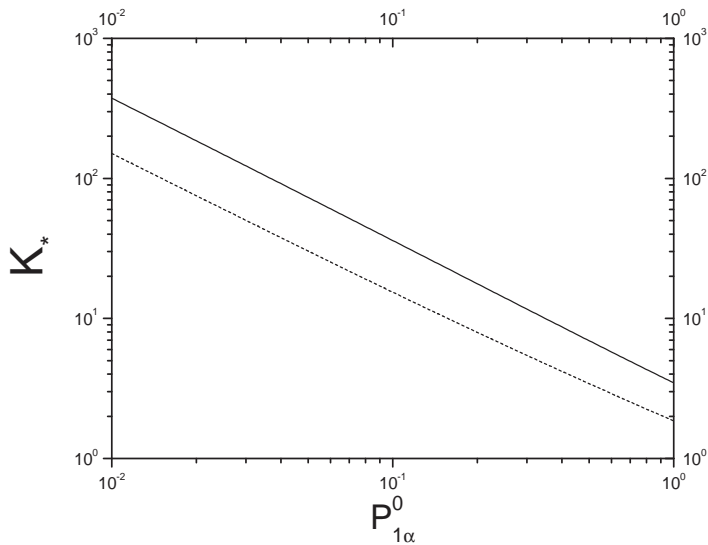,height=7cm,width=11cm}
\caption{Value of $K_{\star}$,
defining the strong wash-out regime for $\d=10\%$ (solid line)
and $\d=50\%$ (dashed line).}
\end{figure}
In Fig. 3 we have plotted the values of the lower bounds
on $M_1$ and $T_{\rm reh}$ for light hierarchical neutrinos,
implying $m_3=m_{\rm atm}$ in the Eq. (\ref{bound}).
These can be obtained plugging
$\xi_1=\sqrt{P^0_{1\a}}\,\k_{1\a}^{\rm f}(K_{1\a})/\k_1^{\rm f}(K_1)\leq 1$
in the  Eq. (\ref{lbM1}).
They correspond to the lowest values in the strong
wash-out regime, when $K_1\geq K_{\star}$.

There are two
possible ways to look at the results. As a function of $P_{1\a}^0$
they get more stringent when $P^0_{1\a}$ decreases so that
the minimum is obtained in the alignment case, corresponding to
the unflavored case. This is clearly visible in the left panel of Fig. 3.
Therefore, one can conclude that
flavor effects cannot help to alleviate the conflict of the leptogenesis
lower bound on the reheating temperature with the gravitino
problem upper bound.
On the other hand, flavor effects can relax the lower bound
for fixed values of $K_1\gg 1$. Indeed, for each value $K_1\gg 1$,
one can choose $P^0_{1\a}=K_{\star}(P^0_{1\a}=1)/K_1$ such that
$K_{1\a}=K_{\star}(P^0_{1\a})$, thus obtaining the highest possible
relaxation in the strong wash-out regime.
The result is shown in the right panel of Fig. 3.
\begin{figure}
\hspace*{-5mm}
\psfig{file=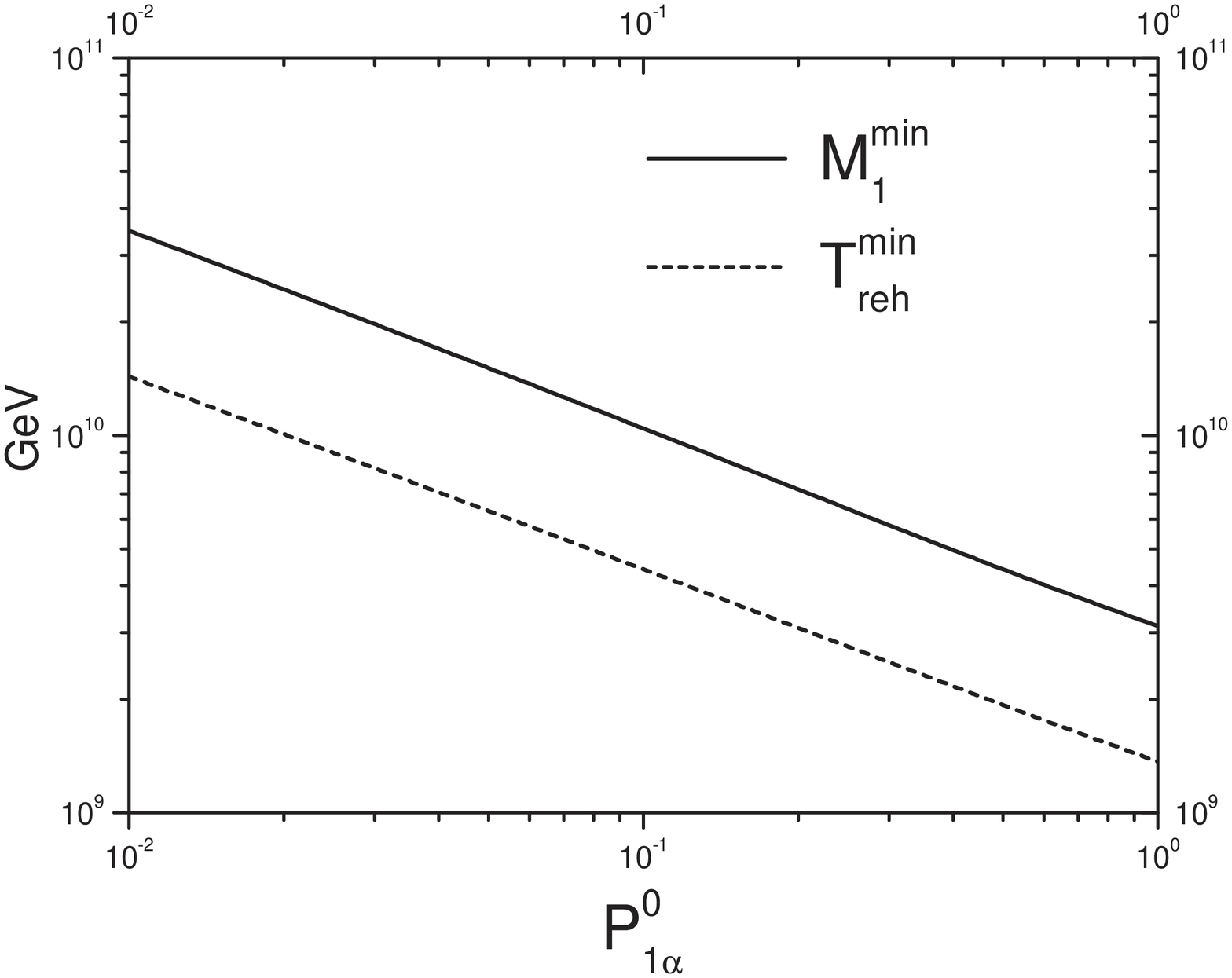,height=8cm,width=9cm}
\hspace{-10mm}
\psfig{file=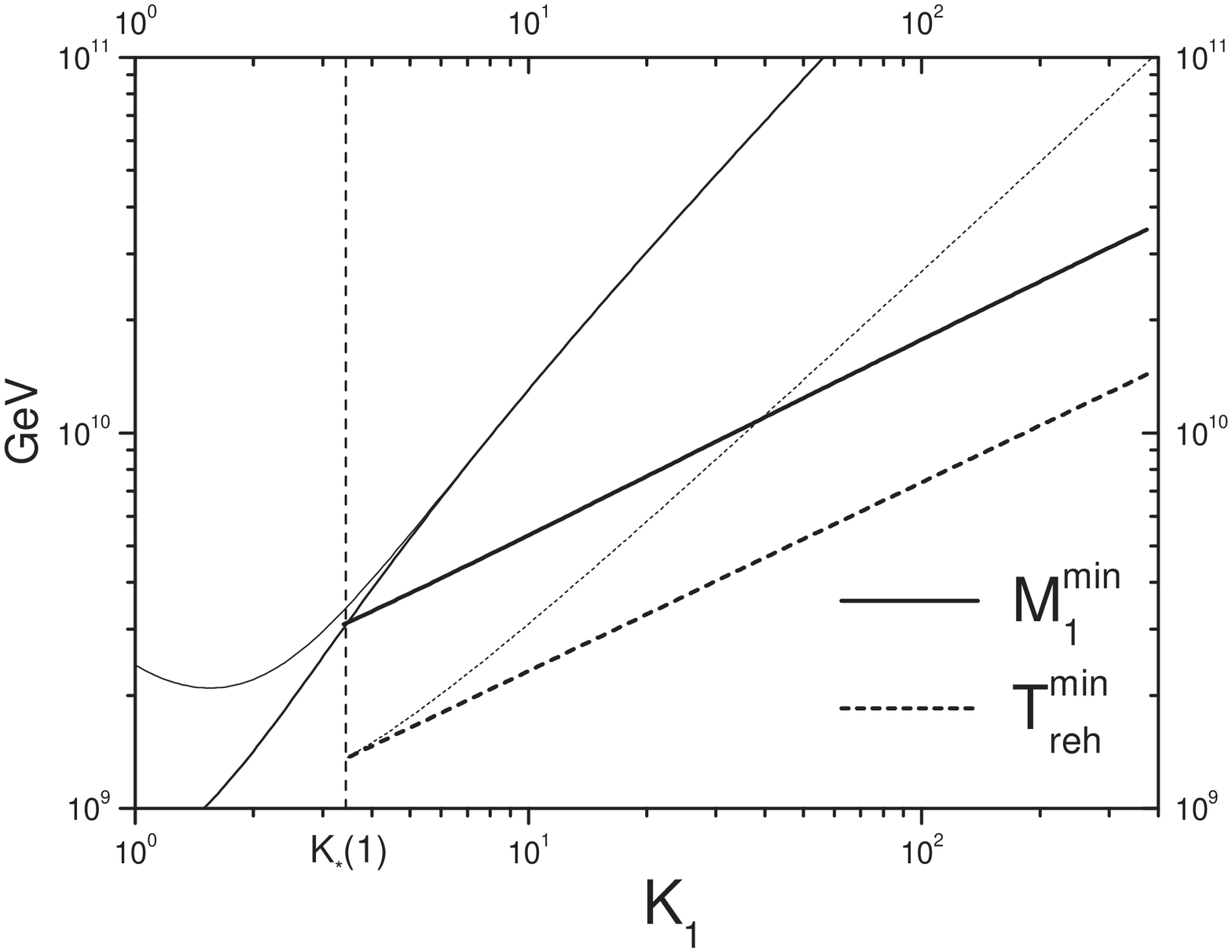,height=8cm,width=9cm}
\caption{Lower bounds on $M_1$ and $T_{\rm reh}$ calculated
choosing $P^0_{1\a}=K_{\star}(P^0_{1\a}=1)/K_1$ such that
$K_1=K_{\star}(P^0_{1\a})$ (thick lines)
and compared with the usual bounds for $P^0_{1\a}=1$ (thin lines). In the left
panel they are plotted as a function of $P^0_{1\a}$, while in the
right panel as a function of $K_1$.}
\end{figure}
It is  important to say that this relaxation is potential.
A direct inspection is indeed necessary to determine
whether it is really possible to achieve at the same time not only
small values of $P^0_{1\a}$, but also a
$\ve_{1\a}$ that is not suppressed compared to $\ve_{1\b\neq\a}$.

Notice that together with the one-flavor dominance case,
one can also envisage, in the three-flavor regime,
a two-flavor dominance case, where
two projectors are equally small, with the third necessarily close
to one, while all the 3 flavored $C\!P$ asymmetries are
comparable.

In the next Section, we will consider a specific example
that will illustrate  what we have discussed on general grounds.
At the same time it will help to understand which are
realistic values for
the projectors and their differences, given a specific set
of see-saw parameters and using the information on neutrino mixing
matrix we have from low-energy experiments.

\section{A specific example}

The previous results have been obtained assuming no restrictions
on the projectors. Moreover, in the one-flavor dominance case,
where there can be a relevant relaxation of the
usual lower bounds holding in the one-flavor approximation,
we have assumed that the upper bound on $\ve_{1\a}$, see Eq. (\ref{bound}),
is saturated independently of the value of the projector.

This assumption does not take into account that the values
of the projectors depend on the different see-saw parameters, in particular
on the neutrino mixing parameters, and that severe
restrictions could apply. Let us show a definite example
considering a particular case for the orthogonal matrix,
\be
\O=R_{13}=
\left(
\begin{array}{ccc}
 \sqrt{1-{\o}^2_{31}}  & 0 &  - {\o}_{31} \\
    0 & 1 & 0 \\
  {\o}_{31} & 0 & \sqrt{1-{\o}^2_{31}}
\end{array}
\right)  \, .
\ee
This case is particularly meaningful,
since it realizes one of the conditions ($\O^2_{21}=0$)
to saturate the bound (\ref{boundtot}) for $\ve_1$ that is given by
$\ve_1 = \overline{\ve}(M_1)\,Y_3\,m_{\rm atm}/(K_1\,m_{\star})$,
with $Y_3\equiv -{\rm Im}[\o^2_{31}]$ \cite{geometry}.
Moreover the decay parameter is given by
\be
K_1=K_{\rm min}\,|1-\o^2_{31}|+K_{\rm atm}\,|\o_{31}^2| \, ,
\ee
where $K_{\rm atm}\equiv m_{\rm atm}/m_{\star}$
and $K_{\rm min}\equiv m_1/m_{\star}$.
The expression (\ref{proj}) for the projector
gets then specialized as
\be
P^0_{1\alpha}= {m_1\,|U_{\a 1}|^2\,|1-\o_{31}^2|
+m_3\,|U_{\a 3}|^2\,|\o_{31}^2| +2\,\sqrt{m_1\,m_3}\,{\rm Re}
[U_{\a 1}\,U_{\a 3}^{\star}\sqrt{1-\o_{31}^2}\,\o_{31}^{\star}]
\over m_1\,|1-\o^2_{31}|+m_3\,|\o_{31}^2|} \, ,
\ee
while, specializing the Eq. (\ref{eps1a}) for $\ve_{1\a}$
and neglecting the term $\propto x_j^{\,-1}$,
one obtains
\begin{eqnarray}\label{full}
r_{1\a} & = &
{Y_3}\,{m_{\rm atm}\over K_1\,m_{\star}}\,\left[
|U_{\a 3}|^2+{m_1^2\over m_{\rm atm}^2}\,
(|U_{\a 3}|^2-|U_{\a 1}|^2)\right] \\ \nonumber
 &-& {m_{\rm atm}\over K_1\,m_{\star}}\,\sqrt{{m_1\over m_{\rm atm}}\,
 {m_3\over m_{\rm atm}}}
 \left[\left({m_1+m_3\over m_{\rm atm}}\right)\,
 {\rm Im}\left[\o_{31}\,\sqrt{1-\o^2_{31}}\right]
 \,{\rm Re}[U^{\star}_{\a1}U_{\a3}]\right. \\ \nonumber
 &  & \;\;\;\;\;\;\;\;\;\;\;\;\;\;\;\;\;\;\;\;\;\;\;\;
 + \left.\left({m_3-m_1\over m_{\rm atm}}\right)\,
 {\rm Re}\left[\o_{31}\,\sqrt{1-\o^2_{31}}\right]
 \,{\rm Im}[U^{\star}_{\a1}U_{\a3}]
\right] \, ,
\end{eqnarray}
where $m_3/m_{\rm atm}=\sqrt{1+m_1^2/m_{\rm atm}^2}$.

If we first consider the case
of fully hierarchical light neutrinos, $m_1=0$, then
\be
P^0_{1\a}={\ve_{1\a}\over \ve_1}=|U_{\a 3}|^2  \,
\hspace{20mm} \mbox{and} \hspace{20mm}
{\D P^0_{1\a}\over 2\,\ve_1}=0  \, .
\ee
Neglecting the effect of the running
of neutrino parameters from high energy to low energy \cite{running},
one can assume that the $U$ matrix can be identified
with the PMNS matrix as measured by neutrino mixing experiments.
We will adopt the parametrization  \cite{PDG}
\begin{equation}
U=\left( \begin{array}{ccc}
c_{12}\,c_{13} & s_{12}\,c_{13} & s_{13}\,e^{-i\,\d} \\
-s_{12}\,c_{23}-c_{12}\,s_{23}\,s_{13}\,e^{i\,\d} &
c_{12}\,c_{23}-s_{12}\,s_{23}\,s_{13}\,e^{i\,\d} & s_{23}\,c_{13} \\
s_{12}\,s_{23}-c_{12}\,c_{23}\,s_{13}\,e^{i\,\d}
& -c_{12}\,s_{23}-s_{12}\,c_{23}\,s_{13}\,e^{i\,\d}  &
c_{23}\,c_{13}
\end{array}\right)
\times {\rm diag(e^{i\,{\Phi_1\over 2}}, e^{i\,{\Phi_2\over 2}}, 1)}
\, ,
\end{equation}
where $s_{ij}\equiv \sin\theta_{ij}$, $c_{ij}\equiv\cos\theta_{ij}$ and
where
we have used $\theta_{13}=0-0.17$, $\theta_{12}=\pi/6$ and $\theta_{23}=\pi/4$.
One then finds
\be
P^0_{1e}\lesssim 0.03 \, ,\hspace{10mm}
P^0_{1\mu}\simeq P^0_{1\tau}\simeq 1/2 \, .
\ee
This result implies a situation where the projector
on the electron flavor is very small and the associated
generated asymmetry as well, while the muon and tauon contributions
are equal. Notice that even though there is
a non-vanishing extra contribution to the muon and tauon
$C\!P$ asymmetries, they have equal absolute value but
opposite sign. Therefore, since the projectors are equal as well,
summing the kinetic equations (\ref{flke}) over $\a$,
one simply recovers the one-flavor
approximation with the wash-out reduced by a factor $2$.
This is a realization of the semi-democratic case that
 we were envisaging at the end of 3.2 where
$K_{\star}\simeq 7$. In this situation flavor effects do not
produce large modifications of the usual results, essentially
a factor 2 reduction of the wash-out in the strong wash-out regime
with a consequent equal relaxation of the lower bounds (see
dashed lines in Fig. 1). Moreover there is practically no difference
between a calculation in the two or in the three-flavor regime.

Let us now consider the effect of a non-vanishing
but small lightest neutrino mass $m_1$,
for example $m_1=0.1\,m_{\rm atm}$. In this case
the results can also depend on the
Majorana and Dirac phases. We will show the results for
$\o^2_{31}$ purely imaginary, the second condition
that maximizes the total $C\!P$ asymmetry if $m_1\ll m_{\rm atm}$
\cite{geometry}. Notice that this is not in general the condition
that maximizes $r_{1\a}$ for non-vanishing $m_1$, however
we have checked that allowing for a real part of $\o_{31}^2$
one obtains similar bounds within a factor ${\cal O}(1)$
if $m_1\ll m_{\rm atm}$.

We have first considered
the case of a real $U$. The results are  only slightly sensitive
to a variation of $\theta_{13}$ within the experimentally
allowed range $0-0.17$. Therefore, we have set $\theta_{13}=0$,
corresponding to $U_{e3}=0$, in all the shown examples.
In the left panel of Fig. 4
\begin{figure}
\hspace*{-5mm}
\psfig{file=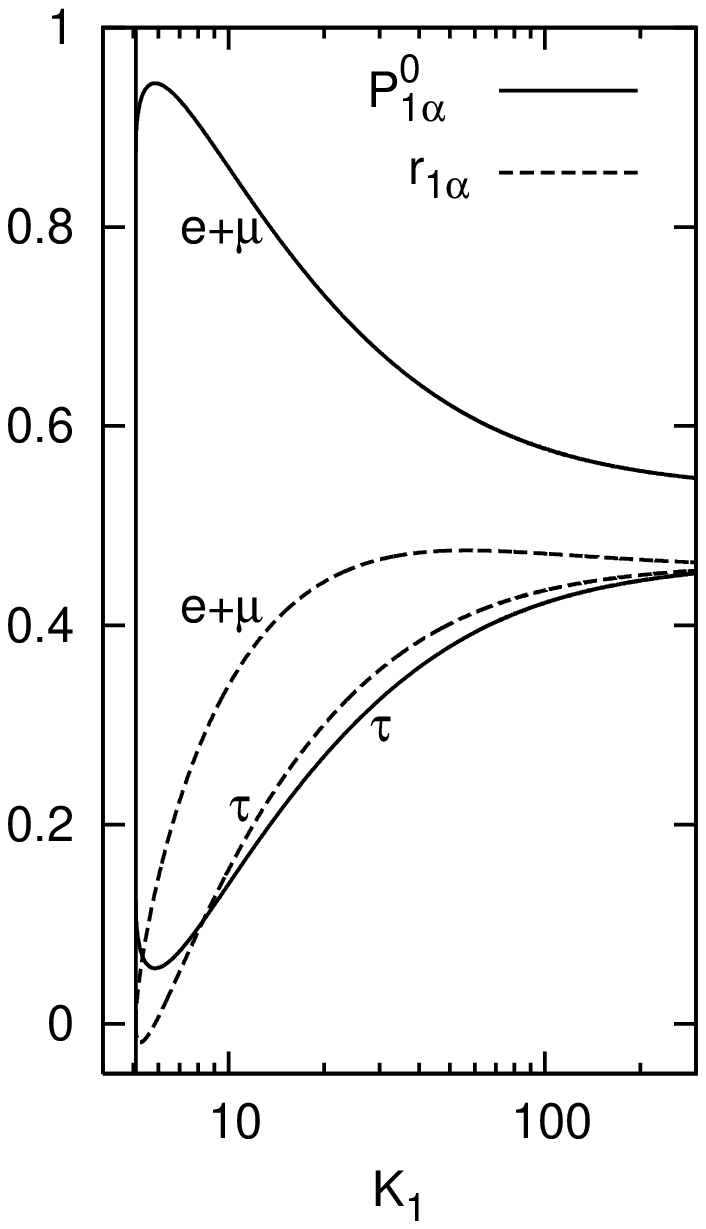,height=7cm,width=53mm}
\hspace{-1mm}
\psfig{file=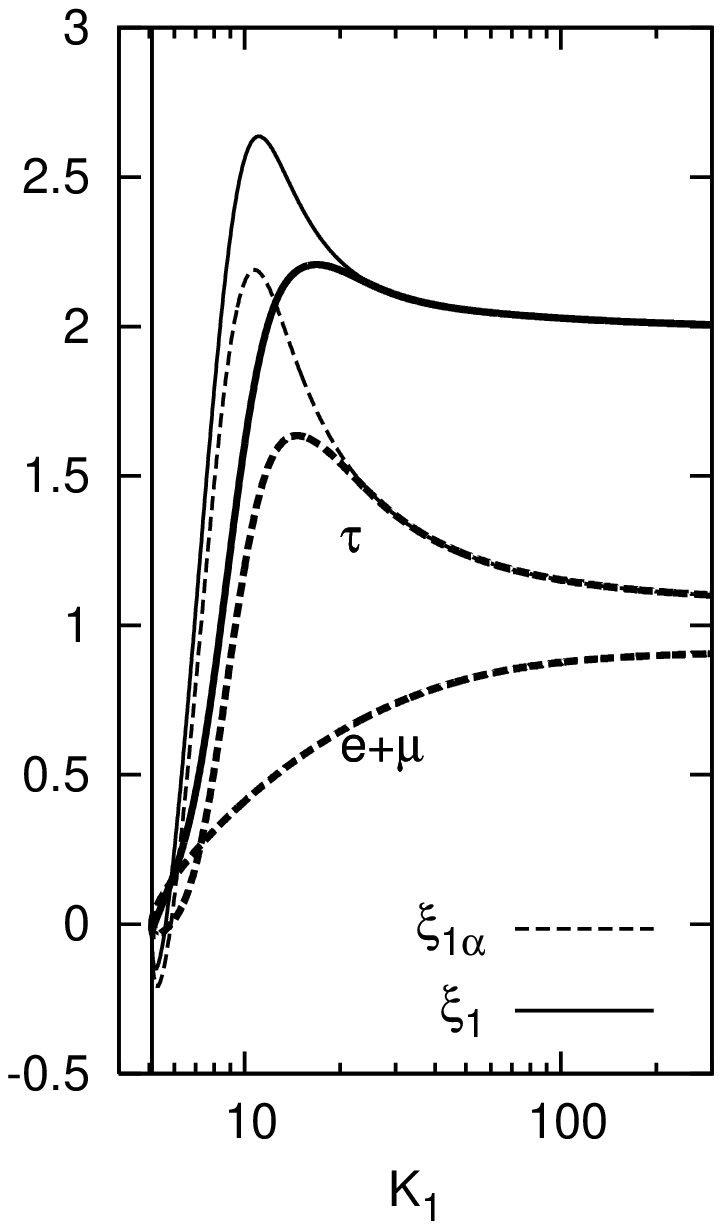,height=7cm,width=53mm}
\hspace{-8mm}
\psfig{file=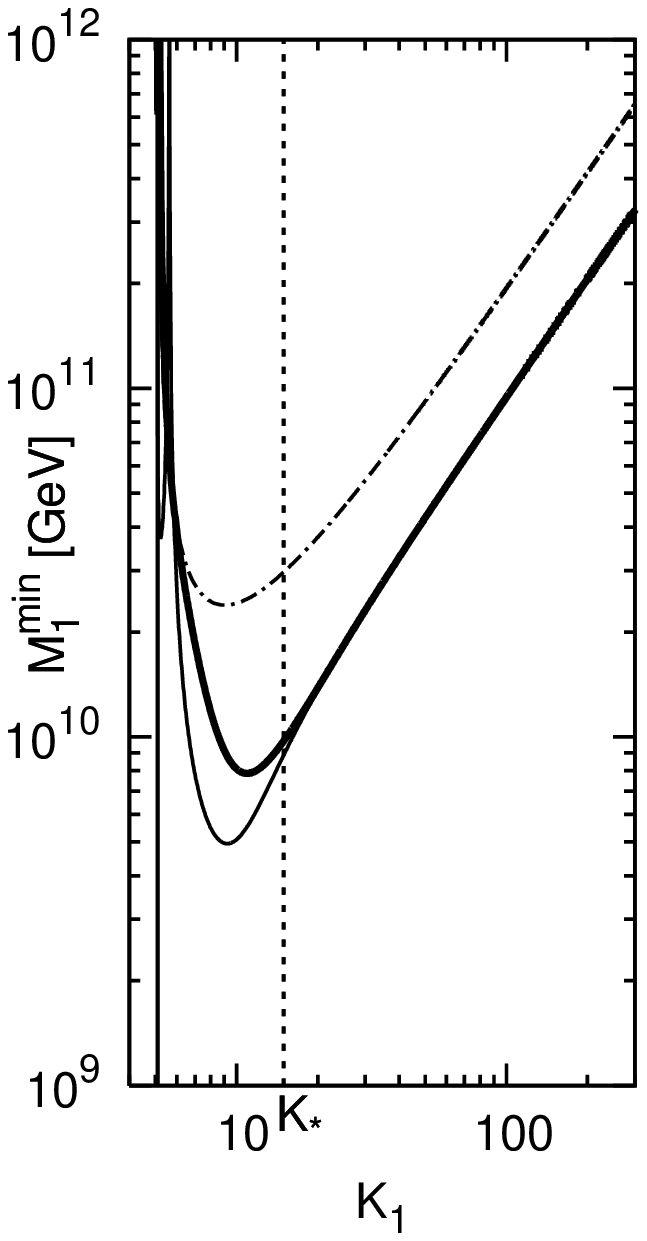,height=7cm,width=62mm}
\caption{Dependence of different quantities on $K_1$ for
$m_1/m_{\rm atm}=0.1$ and real $U$. Left panel: projectors $P^0_{1\a}$
and $r_{1\a}$; central panel: $\xi_{1\a}$ and $\xi_1$ as
defined in Eq. (\ref{xi1a}) for thermal (thin)
and vanishing (thick) initial abundance;
right panel: lower bound on $M_1$ for thermal (thin solid) and
vanishing (thick solid) abundance compared with the one-flavor
approximation result (dash-dotted line).}
\end{figure}
we show the values of the projectors, the $P^0_{1\a}$'s, and
of the $r_{1\a}$'s as a function of $K_1$. The calculations are
performed in the two-flavor regime since we obtain that
successful leptogenesis is possible only for
$M_1 > 10^9\,{\rm GeV}$, where the two-flavor regime applies.
Now a difference between the tauon and the sum of the
muon and electron projectors and asymmetries arises. On the other hand
for $K_1\gg 100$ this difference tends to vanish and
the semi-democratic case is recovered again.

In the central panel we have also plotted the
quantities $\xi_{1\a}$ (cf. (\ref{xi1a})) and their sum
$\xi_1$, where we recall that $\xi_1$ gives the deviation
of the total asymmetry from a calculation in the one-flavor
approximation for hierarchical light neutrinos.
One can see how now
the contribution to the total asymmetry from the tauon flavor
can be twice as large as from the electron plus muon flavor.
For $K_1\gg 100$ the semi-democratic case is restored,
the two contributions tend to be equal
to the one-flavor approximation case and the total
final asymmetry is about twice larger. Finally
in the right panel we have plotted
the lower bound on $M_1$ and
compared them with the results in the one-flavor
approximation (dash-dotted line). At $K_{\star}\simeq 14$
the relaxation is maximum, a factor $\sim 3$.
For $K_1\gg K_{\star}$ the relaxation
is reduced to a factor $2$, as in the semi-democratic case.

Let us now study the effect of switching on phases
in the $U$ matrix, again for $m_1=0.1\,m_{\rm atm}$.
 The most important effect arises from one of the two
 Majorana phases, $\Phi_1$.
In Fig. 5 we have then again plotted, in three panels, the same
quantities as in Fig. 4 for $\Phi_1=-\pi$.
\begin{figure}
\hspace*{-5mm}
\psfig{file=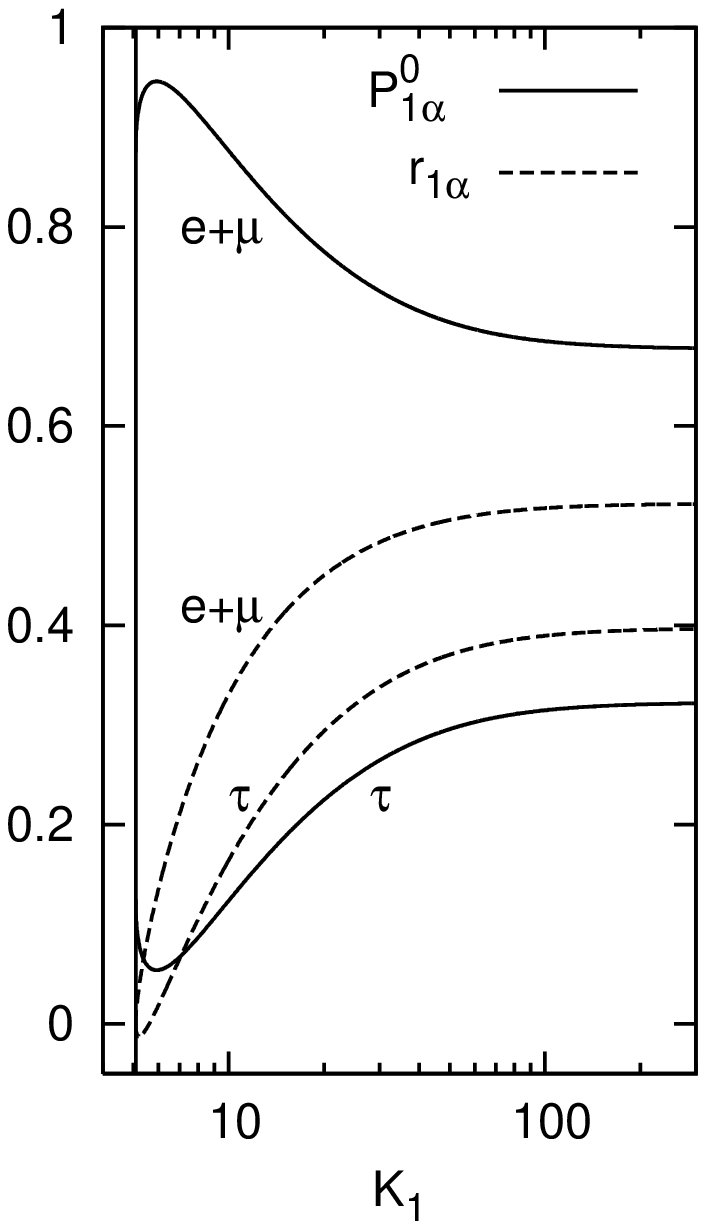,height=7cm,width=53mm}
\hspace{-1mm}
\psfig{file=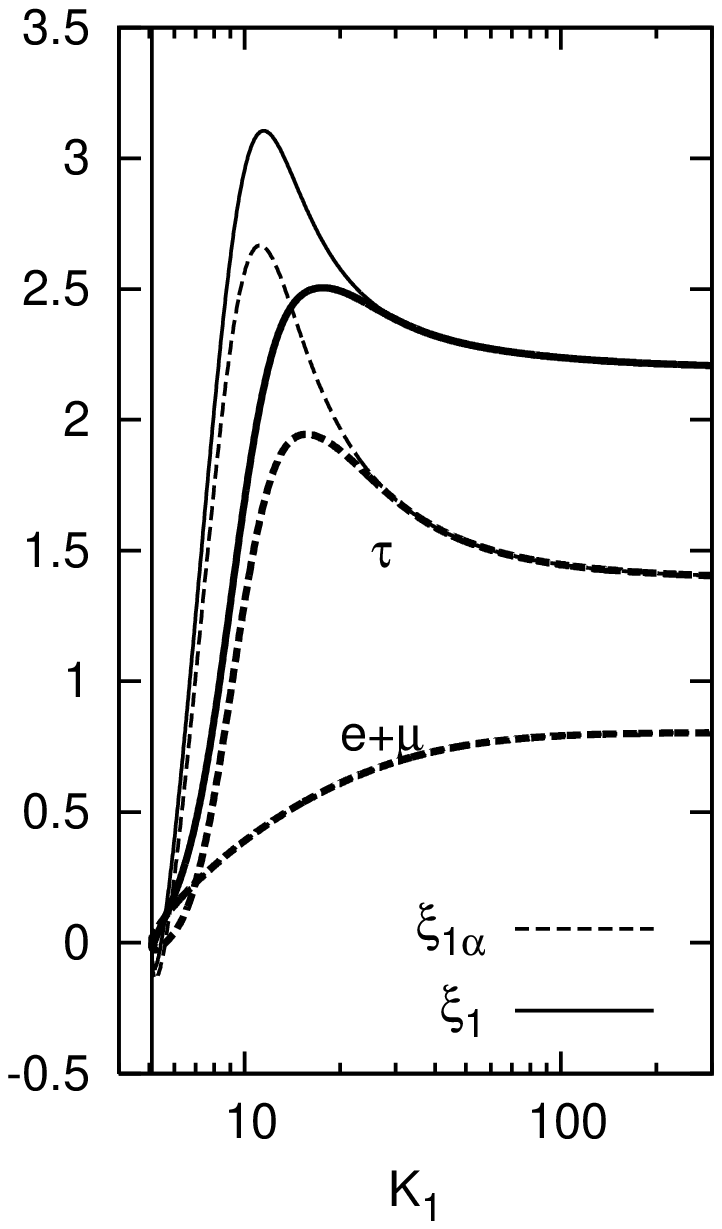,height=7cm,width=53mm}
\hspace{-8mm}
\psfig{file=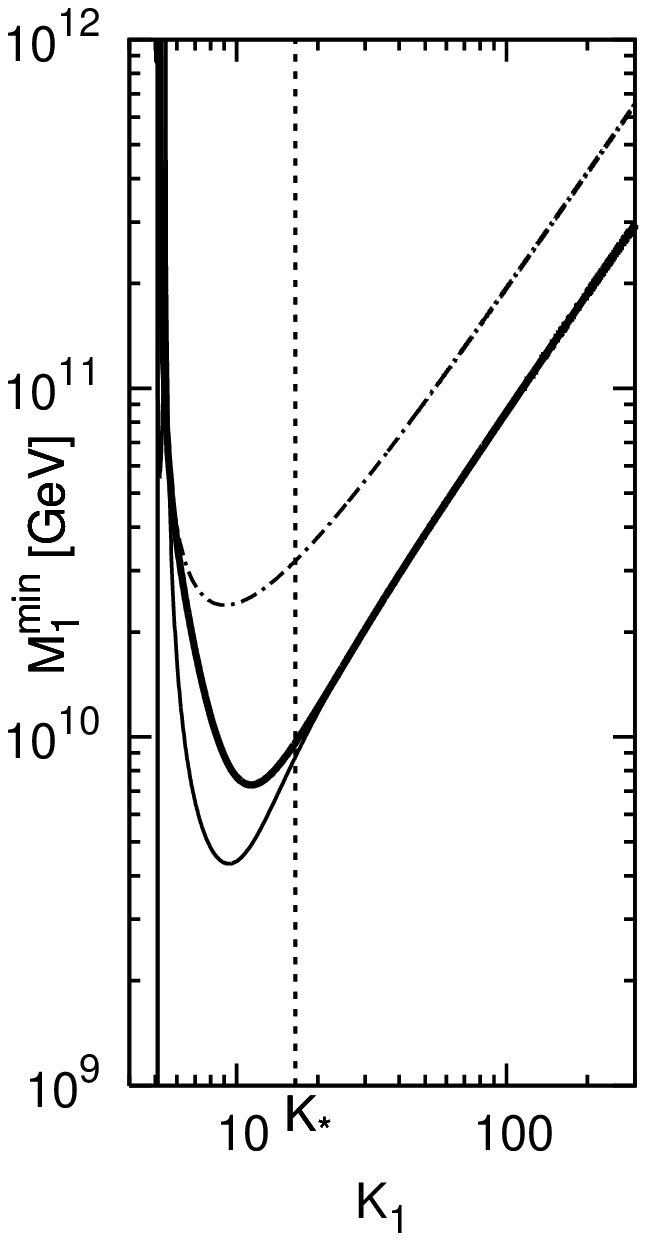,height=7cm,width=62mm}
\caption{As in the previous figure but with one
non vanishing Majorana  phase: $\Phi_1=-\pi$.}
\end{figure}
 One can see how this further increases  the
difference between the $e+\m$ and the $\t$ contributions
and further relaxes the lower bound on $M_1$.
The effect is small for the considered value
$m_1/m_{\rm atm}=0.1$. However, considering
a much larger value while keeping
$\Phi_1=-\pi$, the effect becomes dramatically bigger.
In Fig. 6 we have plotted the same quantities
as in Fig. 4 and Fig. 5 for $m_1/m_{\rm atm}=10$.
\begin{figure}
\hspace*{-5mm}
\psfig{file=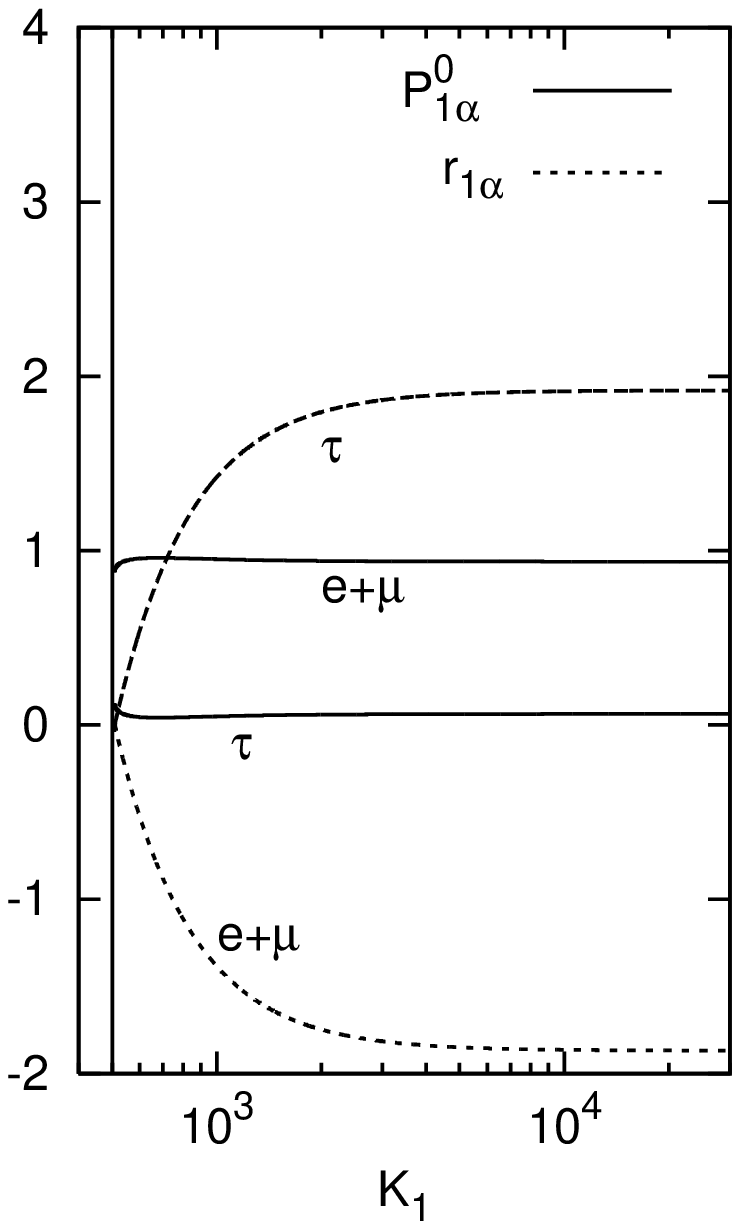,height=7cm,width=53mm}
\hspace{-1mm}
\psfig{file=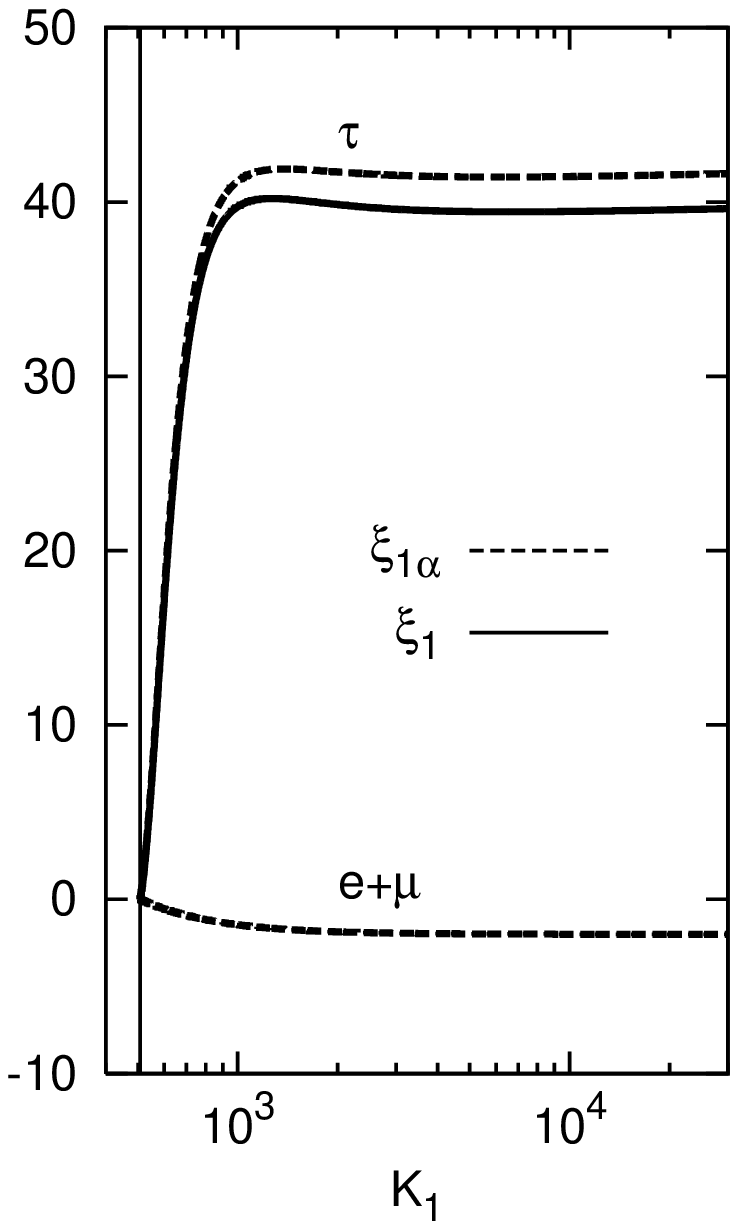,height=7cm,width=53mm}
\hspace{-8mm}
\psfig{file=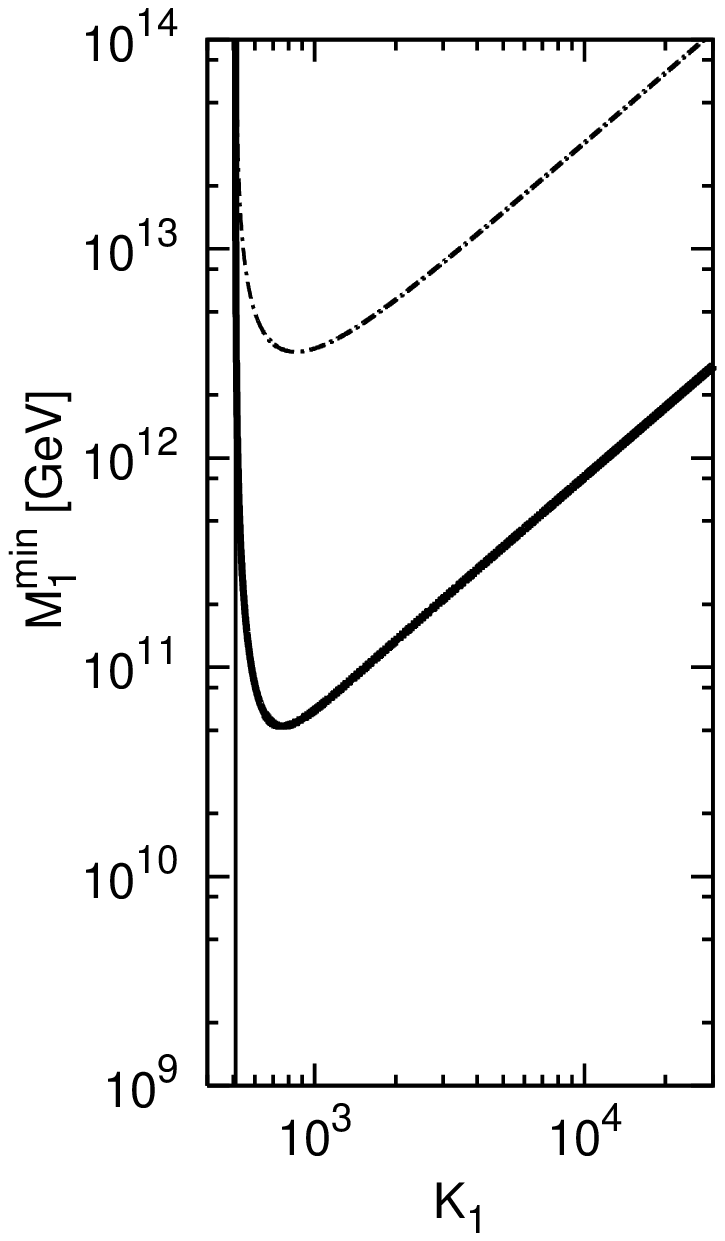,height=7cm,width=62mm}
\caption{Same quantities as in the previous two
figures but with $m_1/m_{\rm atm}=10$
and one non-vanishing Majorana  phase: $\Phi_1=-\pi$.}
\end{figure}
In the left panel one can see that now, for $K_1\gg K_{\rm min}$,
$|r_{1e+\mu}| \simeq |r_{1\t}| \simeq 2$, much larger than
in the previous case. This means that the dominant
contribution to the flavored $C\!P$ asymmetries comes now
from the $\D P_{1\a}$ term. At the same time, very importantly,
$P^0_{1\t}\ll P^0_{1e+\m}$ and in this way,
as one can see in the central panel,
the dominant contribution to $\xi_1$
is given by $\xi_{1\t}$. This case thus finally
realizes a one-flavor dominance case.
The final effect is that the lower bound
on $M_1$ is about one and half order of magnitude relaxed
compared to the one-flavor approximation.

In the left panel of Fig. 7 we have summarized the dependence
on $m_1/m_{\rm atm}$ of the lower bound on $M_1$,
plotting both the case with zero Majorana phase and
the case with $\Phi_1=-\pi$. One can notice how the
the effect of the phase in relaxing the lower bound
increases with $m_1/m_{\rm atm}$.
\begin{figure}
\centerline{
\psfig{file=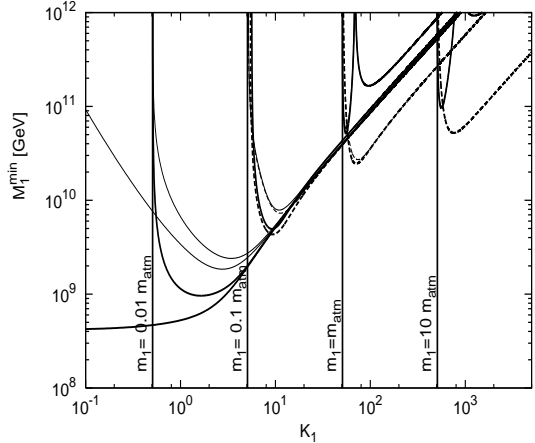,height=75mm,width=6cm,angle=-90}
}
\caption{Lower bound  on $M_1$.
The solid lines are for vanishing phase, while
the dashed lines are for $\Phi_1=-\pi$.
We have shown both the case of vanishing
(thick lines) and of thermal (thin lines)
initial abundance.}
\end{figure}

Finally we wanted to study the interesting case of a
real orthogonal matrix $\O$, implying $\ve_1=0$. If
$m_1=0$ there is practically no produced asymmetry, since the
$\ve_{1\a}\propto \ve_1 = 0$. For a non-vanishing $m_1$ and a
non-real $U$ however the $\D P_{1\a}\neq 0$ and consequently
$\ve_{1\a}\neq 0$.
In Fig. 8 we have shown the results for $m_1/m_{\rm atm}=0.1$
and $\Phi_1=\pi/2$.
We have again plotted the same quantities as in Fig. 4, 5 and 6
in three different panels.
One can see that an asymmetry can be still
produced, as emphasized in \cite{nardi}. However, successful leptogenesis
is possible almost only in the weak wash-out regime. In the
strong wash-out regime, for values $M_1\lesssim 10^{12}\,{\rm GeV}$,
there is a very small allowed region only for
$K_{\star}\simeq 14 \leq K_1 \lesssim 30$. This somehow means that, at least in
the example we are considering and for small $m_1$,
flavor effects give a sub-dominant effect.
\begin{figure}
\hspace*{-5mm}
\psfig{file=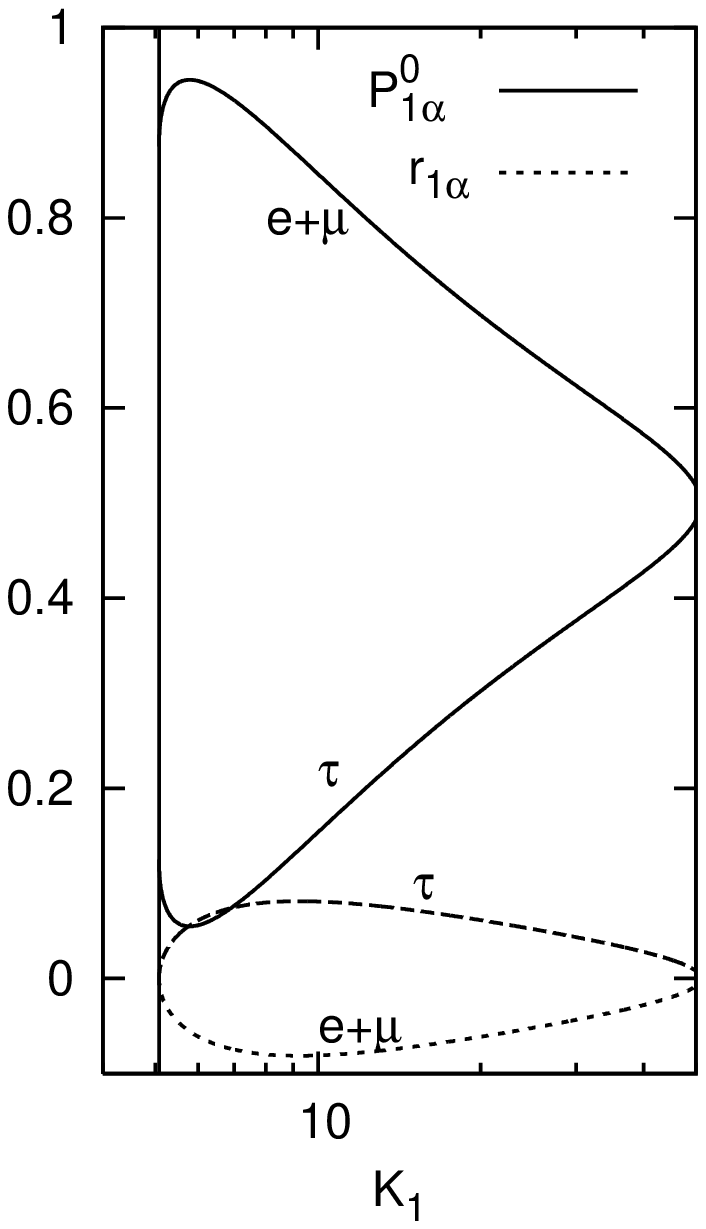,height=7cm,width=53mm}
\hspace{-1mm}
\psfig{file=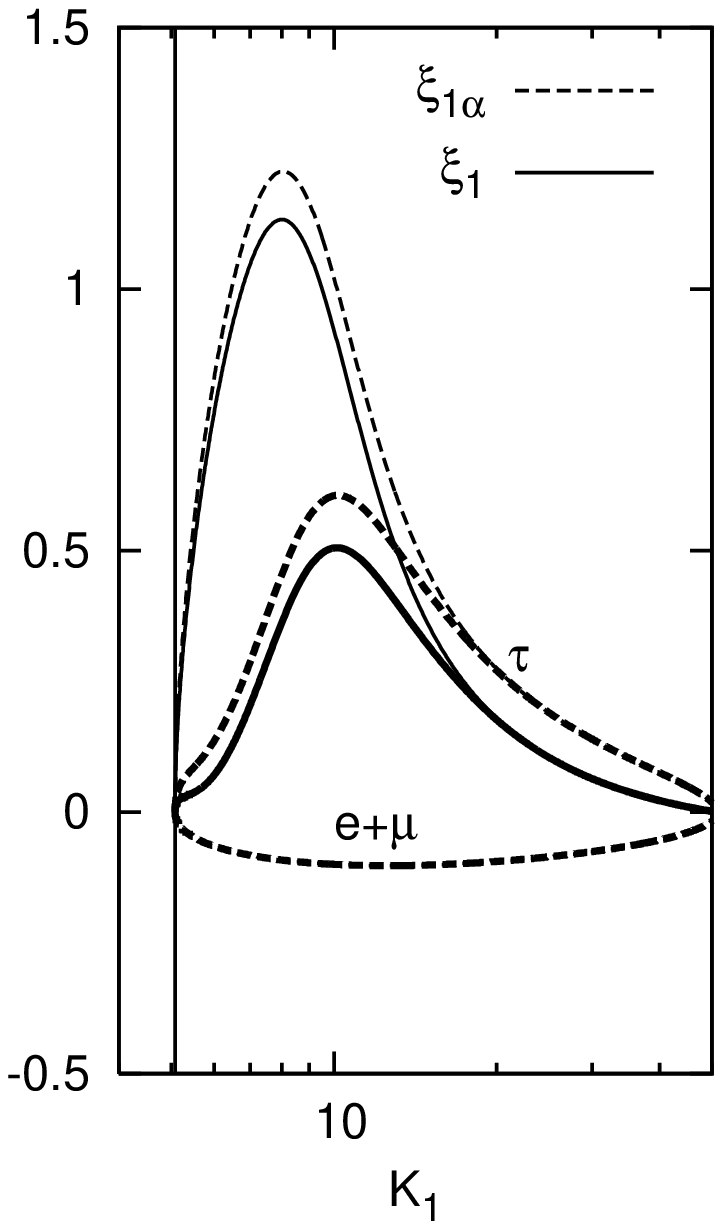,height=7cm,width=53mm}
\hspace{-8mm}
\psfig{file=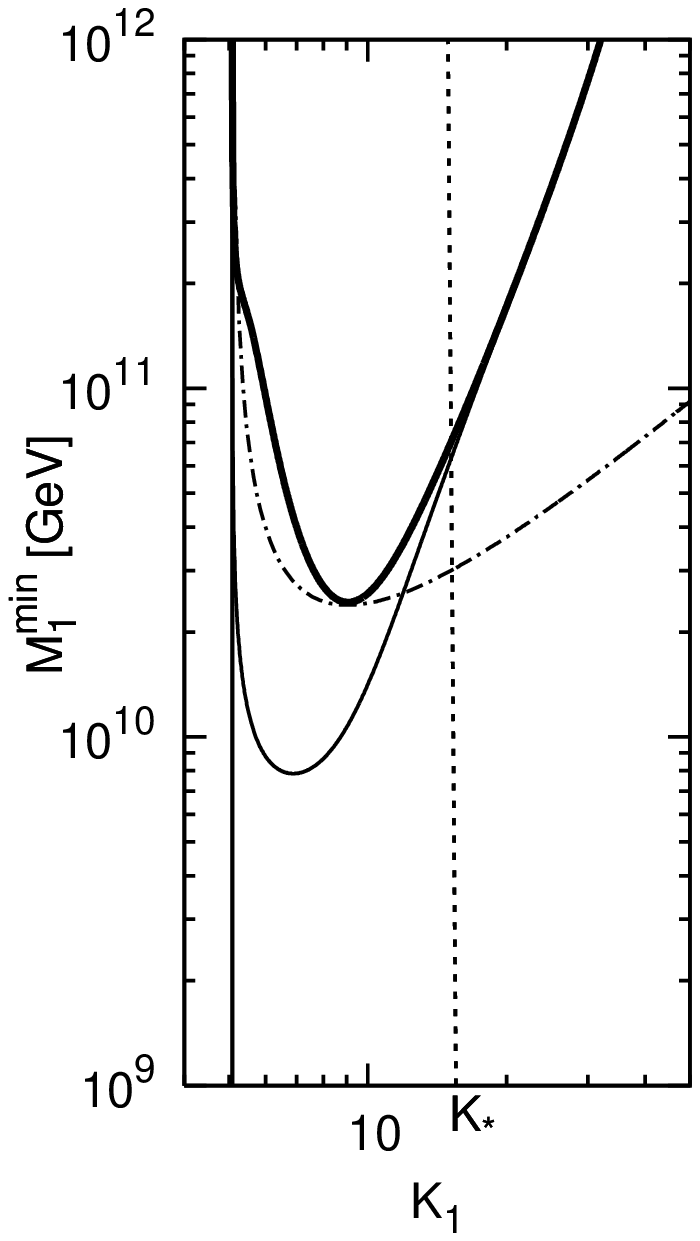,height=7cm,width=62mm}
\caption{Same quantities as in Fig.'s 4,5 and 6 in the case of
real $\O$ for $m_1/m_{\rm atm}=0.1$ and $\Phi_1=\pi/2$. The dot-dashed
still refers to the unflavored case for
$\O=R_{13}$ and purely imaginary $\o^2_{31}$.}
\end{figure}

We can thus draw two remarkable conclusions. The first is that
the relaxation of the lower bounds compared to the one-flavor
approximation is only ${\cal O}(1)$ for $m_1\ll m_{\rm atm}$ but
grows up to one order of magnitude and more for
$m_1\gtrsim 10\,m_{\rm atm}\simeq 0.5\,{\rm eV}$,
in the quasi-degenerate case. Therefore, when
flavor effects are taken into account, one obtains
an opposite result compared to the
one-flavor approximation where one has a suppression
of the final asymmetry for growing absolute neutrino mass
scale, leading to a stringent upper bound $m_1\leq 0.1\,{\rm eV}$
\cite{bound1,bdp2,giudice,annals}. This upper bound seems now to
hold only for $M_1\gtrsim  10^{12}\,{\rm GeV}$ \cite{nardi,abada}.
Notice however that a conclusive answer to this issue
requires a full quantum kinetic calculation describing the evolution
of the density matrices of leptons and anti-leptons \cite{zeno}.
The second is that, when flavor effects are included,
leptogenesis is an interesting example of phenomenology,
beyond $\b\b0\n$, where Majorana phases can play a relevant role.
It must be said however that if the stringent cosmological upper bounds
on the absolute neutrino mass scale will be confirmed,
$m_1\lesssim 0.1-0.2\,{\rm eV}$ \cite{cosmonub}, then
flavor effects can only produce a ${\cal O}(1)$ relaxation of the
lower bound on $M_1$ at large values of $K_1$,
compared to the results from the one-flavor approximation.
This conclusion rigorously applies to the considered example but we
arrived to the same conclusions also in other cases and it seems quite
reasonable that this result holds on more general grounds.
The reason is that for small $m_1\ll m_{\rm atm}$ it does not seem
possible to realize a one-flavor dominated scenario. Further studies are however
needed on this point.

\section{Final discussion}

Flavor effects in leptogenesis can produce
significant deviations from the usual results obtained
within the one-flavor approximation.
However, these are quite restricted in the strong
wash-out regime, imposing that the
final asymmetry is unambiguously determined for different choices of the
initial conditions.

The traditional lowest bounds on $M_1$ and $T_{\rm reh}$ are essentially
the same as in the one-flavor approximation.
Therefore, flavor effects
do not help to solve the conflict with the upper bound on the
reheating temperature coming from the gravitino problem. For this purpose
it is necessary to go beyond a hierarchical
heavy neutrino spectrum \cite{pilaftsis2,bdp2,hambye,blanchet}.

On the other hand, for a fixed value of
the decay parameter $K_1$, the lower bounds can be relaxed
if it is possible to explain the observed asymmetry as dominantly
produced in one flavor. This scenario seems not easily achievable since
typically for low values of the projector, the final associated asymmetry is
also lower.

An interesting feature of flavor effects is that they depend
in general on the low energy parameters \cite{endo,nardi},
in particular on the lightest neutrino mass $m_1$
and on the phases in the neutrino mixing parameters. We have studied
an interesting example assuming the conditions for maximal total
$C\!P$ asymmetry. We showed that
for vanishing phases and lightest neutrino mass one essentially
recovers the one-flavor approximation just with halved wash-out,
the semi-democratic case, where the electron flavor gives a
negligible contribution while the final asymmetry is
equally shared between the muon and the tauon flavor contributions.
However, switching on simultaneously a non-vanishing
lightest neutrino mass and Majorana phases, lead to an enhancement of
the $C\!P$ asymmetry in the tauon flavor. In this way
a tauon-flavor dominated picture is realized, where
the wash-out is much lower compared to the case when
flavor effects are neglected. It is then quite interesting that there is an
emerging interplay  between future low energy experiments and
flavored leptogenesis predictions on the matter-antimatter
asymmetry of the Universe.

\vspace{2mm}
\noindent
\textbf{Acknowledgments}\\
It is a pleasure to thank  M.~Pl\"{u}macher and G.~Raffelt
for useful comments and discussions.
\noindent

\vspace{2mm}
\noindent
\textbf{Note added}\\
During the revision of the paper, the work \cite{zeno} was completed,
finding a more restrictive condition for the importance of flavor effects and
for the validity of the Eq.'s (\ref{flke}). The condition
is particularly restrictive in the case of one-flavor dominance,
relevant for the results presented in Figures 6 and 7 and for the possibility
of circumventing the upper bound on the neutrino masses holding
in the unflavored case. We should then warn the reader that these results
need to be confirmed by a more general quantum kinetic analysis.
\noindent

\hspace{15mm}
\section*{Appendix}

In this appendix we discuss the effects of $\D L=1$ scatterings
\cite{many,many2,giudice} on the determination of the efficiency
factors $\k_{1\a}^{\rm f}$ and more particularly on $K_{\star}$.
It has been recently noticed that scatterings also
contribute to the $C\!P$ violating source term with the
same $C\!P$ asymmetries $\ve_{i\a}$ associated to inverse decays
\cite{pilaftsis,nardi2,abada2}.
Therefore, here we also want to discuss the effect of this term
extending the analytic procedure
of \cite{annals}, where it was not included. Including scatterings the
kinetic equations (\ref{flke}) in the $N_1$DS get modified as
\bea\label{flavors}
{dN_{N_1}\over dz} & = & -(D_1+S_1)\,(N_{N_1}-N_{N_1}^{\rm eq}) \\
{dN_{\D_{\a}}\over dz} & = &
\ve_{1\a}\,(D_1+S_1)\,(N_{N_1}-N_{N_1}^{\rm eq})
-P_{1\a}^{0}\,\,(W_1^{\rm ID}+W_1^{\rm S})\,N_{\D_{\a}} \, .
\eea
The notation is the same as in \cite{annals},
except that here we are keeping the index referring to
the lightest RH neutrino. Defining
\be
j_1(z)\equiv 1+{S_1\over D_1} \, ,
\ee
the expression (\ref{k1a}) for  $\k_{1\a}^{\rm f}$ in
the case of initial thermal abundance
($N_{N_1}^{\rm in}=1$) gets generalized as
\be\label{k1aj}
\k(K_{1\a}) \equiv
{2\over z_B(K_{1\a})\,K_{1\a}\,j_1(z_B)}\,\left(1-e^{-{K_{1\a}\,
z_B(K_{1\a})\,j_1(z_B)\over 2}}\right) \, .
\ee
Comparing with the result obtained when scatterings are neglected
in the $C\!P$ violating term, there is now $j_1(z_B)$ instead of
$j_1^2(z_B)$, making the effect of scatterings even less important
than it was already. This could be relevant in the case
of a dynamically generated abundance in the weak wash-out regime,
since scatterings are able to produce a larger abundance at
the decay time. However again, when the effect of scatterings is
included in the $C\!P$ violating term as well, this enhancing effect
is greatly reduced. The Eq. (\ref{k-}) for $\k_{-}^{\rm f}$,
when scatterings are included, becomes indeed
\be
\k_{-}^{\rm f}(K_1,P_{1\a}^{0})\simeq
-{2\over P_{1\a}^{0}}\ e^{-{N_{S}(K_{1\a})\over 2}}
\left(e^{{P_{1\a}^{0}\over 2}\,\overline{N}_S(K_1)} - 1 \right) \, ,
\ee
where
\begin{equation}\label{nkall}
N_S(K_{1\a}) = {3\,\p K_{1\a}\over 4}+
\int_{0}^{\infty}\,dz'\,W_{1\a}^{\rm ID}\,{S_1\over D_1} .
\end{equation}
The expression (\ref{k+}) for $\k_{+}^{\rm f}$
remains formally unchanged,
but now $\overline{N}_S(K_1)$ has to be calculated
replacing $N(K_1)$ with $N_S(K_1)$ into the Eq. (\ref{nka}).
Even without specifying at all the scattering term $S_1$, there is
quite an interesting general feature resulting from the inclusion
of scatterings in the $C\!P$ violating term. One can indeed
calculate the behavior of $\k_{1\a}^{\rm f}=\k_{+}^{\rm f}+\k_{-}^{\rm f}$
for $K_1\rightarrow 0$ discovering that \cite{abada2}
\be\label{K2}
\k_{1\a} \rightarrow P_{1\a}^0\,
\left[{N_{N_1}(z_{\rm eq})\over 2}\right]^2 \propto K_1^2 \, ,
\ee
exactly as in the case when scatterings are neglected \cite{annals},
except for the modified $N_{N_1}(z_{\rm eq})$.
The reason is that now there is a proportionality between
asymmetry generation and neutrino production both in
scatterings and in inverse decays. Conversely,
when scatterings are neglected in the $C\!P$ violating term,
there is an overproduction of neutrinos due to scatterings not
compensated by an equal (negative) production of asymmetry, such that
scatterings enhance the final asymmetry that is $\propto K_1$ instead
of $K_1^{\,2}$.
Notice however that in the strong wash-out regime there is practically no
difference.

In Fig. 9 we have plotted the final efficiency factor $\k_{1\a}^{\rm f}$
using a useful analytic expression for $j_1(z)$ given by \cite{annals}
\be
j_1(z)\simeq
\left[{z\over a}\,\ln\left(1+{a\over z}\right)+{K_1^S\over K_1\,z}\right]\,
\left(1+{15\over 8\,z}\right) \, ,
\ee
where $a={K_1/K_1^S\,\ln(M_1/M_h)}$ and $K_1^S\simeq 0.1\,K_1$.
\begin{figure}
\hspace*{-5mm}
\psfig{file=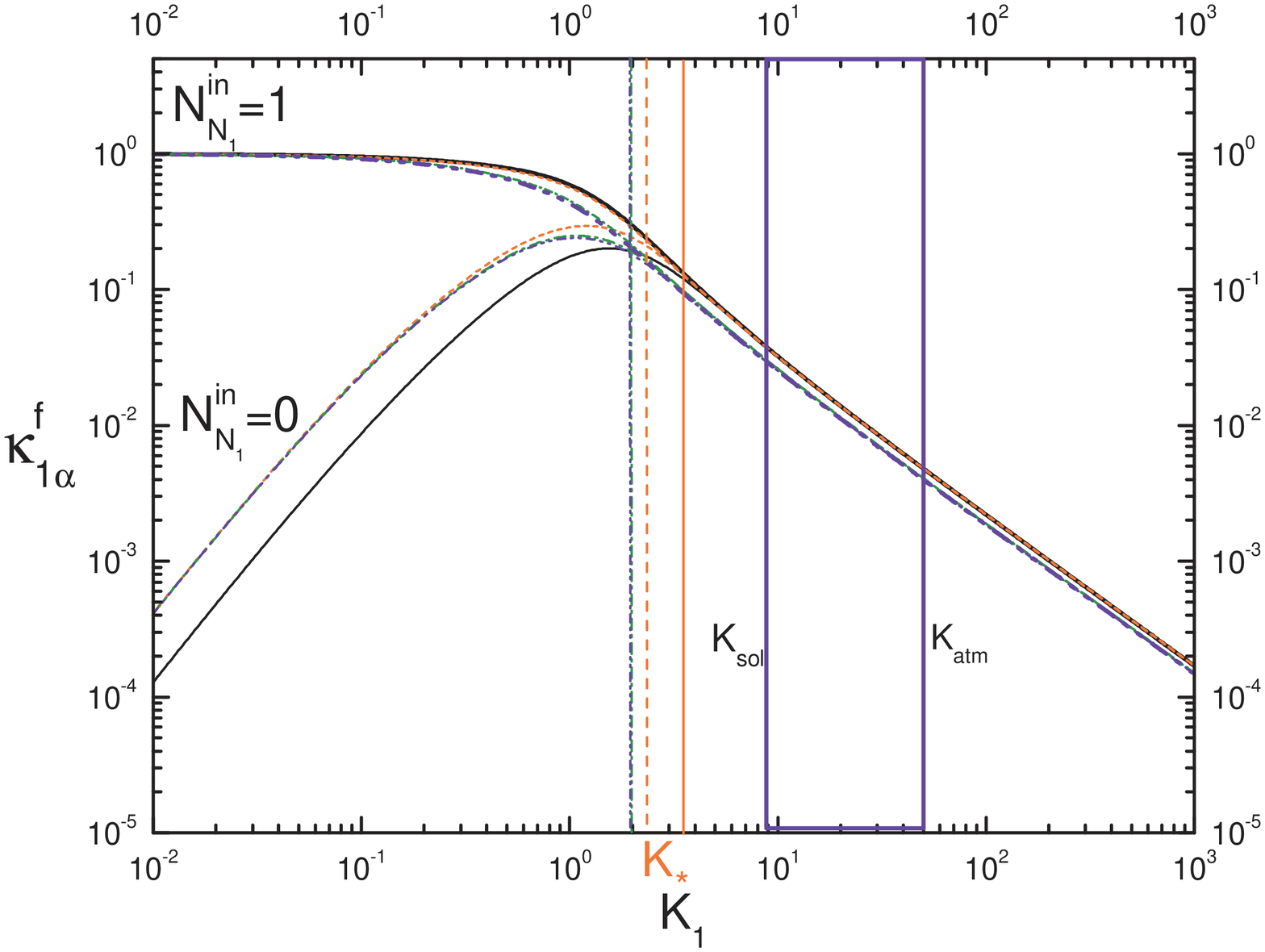,height=8cm,width=9cm}
\hspace{-10mm}
\psfig{file=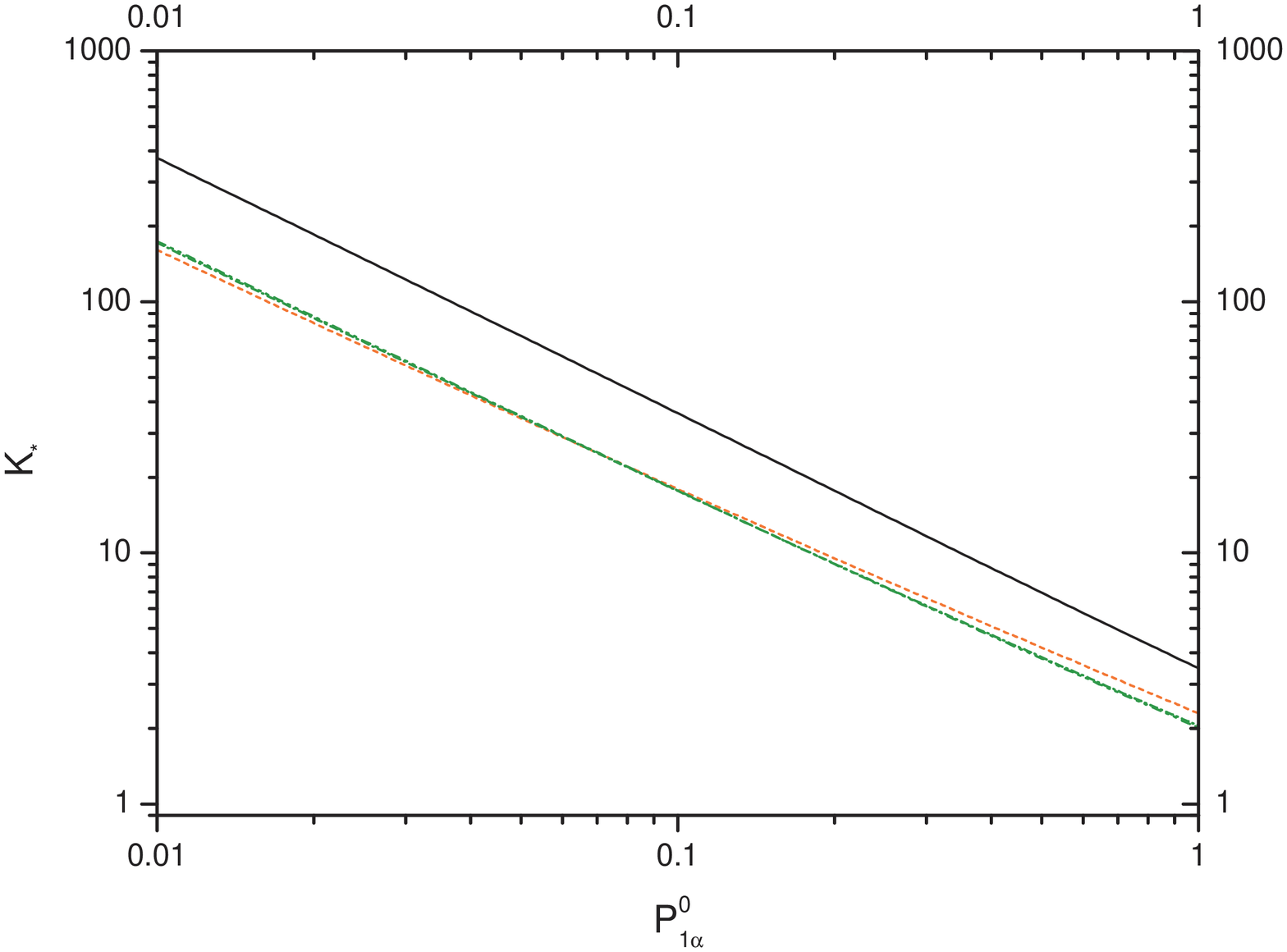,height=8cm,width=9cm}
\caption{Efficiency factor (left panel) and $K_{\star}$ (right panel)
when scatterings are taken into account for 3 values
of $M_1/M_{\rm H}$: $10$ (dash), $10^5$ (dot-dash), $10^{10}$ (dot-dot-dash),
to be compared with the case when they are neglected (solid lines).}
\end{figure}
This approximation works well when scatterings involving gauge
bosons and thermal effects \cite{many2,giudice} are neglected.
The different curves are obtained for $M_1/M_H=10,10^5$ and $10^{10}$
and compared with the case when scatterings are neglected
(solid lines). Notice that the chosen  values for $M_1/M_H$
are the same as in \cite{annals}. A comparison shows
that the results exhibit now a greatly reduced
model dependence, here illustrated
by the different values of $M_1/M_H$,
also in the weak wash-out regime.
Including thermal effects and scatterings
involving gauge bosons would yield similar  results.

In conclusion, the inclusion of scatterings in the
$C\!P$ violating terms greatly reduces the theoretical uncertainties
in the weak wash-out regime, even though these are still much
larger than in the strong wash-out regime. In particular the
asymptotic behavior $\propto K^2$ at small $K$ is a robust
feature, as explained by the Eq. (\ref{K2}).
In the right panel we have shown the corresponding value of $K_{\star}$.
Notice that thermal effects are approximately reproduced by
a thermal Higgs mass $M_H\simeq 0.4\, T$ \cite{many2,giudice} and in the
strong wash-out regime this is approximately equivalent to the case
$M_1/M_H\simeq 10$ where scatterings produce just a correction in the
calculation of $K_{\star}$. Even for larger $M_1/M_H$ the
value of $K_{\star}$ calculated neglecting scatterings
is within the precision needed for the results of the paper,
thus justifying the approximation to neglect scatterings.

\end{document}